\documentclass[prd,nofootinbib,twocolumn,preprintnumbers]{revtex4}

\pdfoutput=1

\usepackage{hyperref}

\usepackage{color}
\usepackage[dvipsnames]{xcolor} 
\usepackage[utf8]{inputenc}
\usepackage[squaren]{SIunits}
\usepackage{amsmath}
\usepackage{amsfonts}
\usepackage{slashed}
\usepackage{graphicx}
\usepackage{subfigure}
\usepackage{rotating}

\usepackage{graphicx,amsmath,amssymb,epsfig,verbatim,mathrsfs,array,layout,textcomp,latexsym}
\usepackage{multirow}

\allowdisplaybreaks[1]

\newcommand{\TeV}{\,\mathrm{TeV}}
\newcommand{\GeV}{\,\mathrm{GeV}}

\newcommand{\bra}[1]{\left\langle{#1}\right\vert}
\newcommand{\ket}[1]{\left\vert{#1}\right\rangle}
\def\ubar{\overline{u}}

\def\lbar{\overline{\ell}}

\begin{document}

\setlength{\parindent}{0pt}

\preprint{DO-TH 20/03}

\title{Exploiting CP--asymmetries in rare charm decays}

\author{Rigo Bause}
\email{rigo.bause@tu-dortmund.de}
\author{Hector Gisbert}
\email{hector.gisbert@tu-dortmund.de}
\author{Marcel Golz}
\email{marcel.golz@tu-dortmund.de}
\author{Gudrun Hiller}
\email{ghiller@physik.uni-dortmund.de}
\affiliation{Fakultät für Physik, TU Dortmund, Otto-Hahn-Str.\,4, D-44221 Dortmund, Germany}

\begin{abstract} 
We analyze patterns from CP--violating new physics (NP) in hadronic and semileptonic rare charm  $\vert \Delta c \vert=\vert \Delta u\vert=1$ transitions. 
Observation of direct CP--violation in hadronic decays, as in $\Delta A_{\text{CP}}$, provides opportunities for  $c \to u \,\ell^+ \ell^-$, $\ell=e,\mu$  transitions, and vice versa.
For the concrete case of flavorful, anomaly-free $Z^\prime$--models a NP--interpretation of $\Delta A_{\text{CP}}$ suggests measurable CP--asymmetries  in semileptonic decays such as $D \to \pi\, \ell^+ \ell^-$ or $D \to \pi \pi\, \ell^+ \ell^-$.
Conversely, an observation of CP--violation in  $c \to u\, e^+ e^-$ or  $c \to u\, \mu^+ \mu^-$  decays supports a NP--interpretation of $\Delta A_{\text{CP}}$.
Flavorful $U(1)^\prime$--extensions provide explicit U--spin and isospin breaking which can be probed in patterns of hadronic decays
of charm mesons. We work out signatures for CP--asymmetries in $D^0 \to \pi^+ \pi^-$,  $D^0 \to K^+ K^-$ and  $D^0 \to \pi^0 \pi^0$, $D^+ \to \pi^+ \pi^0$ decays, which can be probed in the future at LHCb and Belle II and provide further informative cross checks.
\end{abstract}

\maketitle

\section{Introduction}

Suppressions of standard model (SM)  amplitudes due to accidental symmetries provide  useful directions for  searches for new physics (NP).
Among the salient features of $\vert \Delta c \vert=\vert \Delta u\vert=1$
transitions within the SM are a strong Glashow-Iliopoulos-Maiani (GIM)--suppression and small CP--violation. 
Hierarchies of the Cabibbo-Kobayashi-Maskawa (CKM) matrix $V$ suggest SM CP--violation at the order of $\text{Im}( V_{cb}^* V_{ub}/(V_{cs}^* V_{us})) \sim 7  \cdot 10^{-4}$,
somewhat below LHCb's observation of CP--violation in charm  \cite{Aaij:2019kcg}
 \begin{align}
 \begin{split}
        \Delta A_{\text{CP}}&= A_{\text{CP}}(K^+K^-)-A_{\text{CP}}(\pi^+\pi^-)\,\\&=\,(-15.4\pm 2.9) \cdot  10^{-4}~,\end{split}\label{eq:deltaACP}
    \end{align}
where
    \begin{align}
        A_{\text{CP}}(f)=\frac{\Gamma(D^0\to f)-\Gamma(\overline{D}^0\to f)}{\Gamma(D^0\to f)+\Gamma(\overline{D}^0\to f)}~, \label{eq:acpdef}
    \end{align}
and the corresponding world average~\cite{Amhis:2019ckw}
\begin{align}
    \label{eq:worldaverage}
    \Delta A_{\text{CP}}^{\text{HFLAV}}=(-16.4\pm2.8) \cdot 10^{-4} \, . 
\end{align}
While this leaves room for NP, due to the sizable uncertainties of hadronic  $D$--decays,  Eqs.~\eqref{eq:deltaACP} and \eqref{eq:worldaverage} provide no clear-cut sign of NP. On the other hand, $\Delta A_{\text{CP}}$ as large as the permille level is non-trivial to achieve in concrete models of NP.
Correlations with other observables in charm and the down-quark sector exist, which are subject to
partly very strong flavor constraints.
For recent works, see Refs.~\cite{Dery:2019ysp, Chala:2019fdb,Buccella:2019kpn,Li:2019hho,Soni:2019xko,Cheng:2019ggx,Khodjamirian:2017zdu,Kagan:2020vri,Nierste:2020eqb,Pich:2019pzg}.
Turning this around, the study of patterns using  different sectors can hence disfavor or support a particular $\Delta A_{\text{CP}}$ interpretation, and vice versa.
 
In this work we pursue a global analysis of CP--asymmetries in rare hadronic and semileptonic charm decays.
Our focus is on NP patterns induced by four-fermion operators. Links via dipole operators between hadronic and semileptonic CP--asymmetries in $D\to \pi\, \ell^+ \ell^-$ decays have been pointed out by Ref.~\cite{Fajfer:2012nr}.
We work out predictions and correlations for anomaly-free $Z^\prime$--extensions of the SM with generation-dependent $U(1)^\prime$--charges, 
see Refs.~\cite{Ellis:2017nrp,Allanach:2018vjg,Rathsman:2019wyk,Costa:2019zzy,Bause:2019vpr,Aebischer:2019blw,Choudhury:2020cpm} for recent phenomenological works.
Flavorful charges can give rise to explicit isospin and U--spin breaking effects.
It is our goal  to work out corresponding experimental signatures for  hadronic charm decays, exploiting yet another SM null test strategy  in charm \cite{deBoer:2018buv}.
 
This  paper is organized as follows: In Section \ref{sec:CPX} we  briefly review CP--violation in hadronic $D$--decays, $D$--mixing and semileptonic $c \to u \,\ell^+ \ell^-$ transitions.
In Section \ref{sec:Zp} we analyze effects of anomaly-free $U(1)^\prime$--extensions with generation-dependent charges in 
hadronic 2-body $D$--decays and how $D$--mixing constraints can be evaded to address $  \Delta A_{\text{CP}}$.
Patterns among CP--asymmetries in $D^0 \to \pi^+ \pi^-$,  $D^0 \to K^+ K^-$, $D^0 \to \pi^0 \pi^0$ and $D^+ \to \pi^+ \pi^0$ decays are worked out in Section \ref{sec:patterns}.
Correlations with CP--asymmetries in rare semileptonic decays are studied in Section~\ref{sec:slCP}. We conclude in Section~\ref{sec:con}. Auxiliary information is given in several appendices.
 
\section{CP--phenomenology in charm}\label{sec:CPX}

We review CP--violation in hadronic $D$--decays (Section~\ref{sec:2body}), $D$--mixing (Section~\ref{sec:mixing}) and semileptonic $c \to u\, \ell^+ \ell^-$ processes (Section~\ref{sec:sl}).

\subsection{Direct CP--violation in \texorpdfstring{$\boldsymbol{D^0\to\pi^+\pi^-,\,K^+K^-}$}{Dto Pi+Pi-orK+K-}}\label{sec:2body}

The single-Cabibbo-suppressed (SCS) $D^0$($\overline{D}^0$) decay amplitudes $\mathcal{A}_f$ ($\overline{\mathcal{A}_f}$) to CP--eigenstates $f$ can be written as
    \begin{align}
    \begin{split}
    \mathcal{A}_f\,&=\,\mathcal{A}_f^T\,\text{e}^{\text{i}\,\phi_f^T}\left[1+r_f\,\text{e}^{\text{i}\,(\delta_f+\phi_f)}\right]~,\\
    \overline{\mathcal{A}_f}\,&=\,\eta_{\text{CP}}\,\mathcal{A}_f^T\,\text{e}^{-\text{i}\,\phi_f^T}\left[1+r_f\,\text{e}^{\text{i}\,(\delta_f-\phi_f)}\right]~,
    \end{split}\label{eq:Af}
    \end{align}
where $\eta_{\text{CP}}=\pm 1$ is the CP--eigenvalue of $f$. The dominant SCS ``tree'' amplitude in the SM is denoted by $\mathcal{A}_f^T\,\text{e}^{\pm\,\text{i}\,\phi_f^T}$, and $r_f$ parametrizes the relative magnitude of all subleading amplitudes. Inserting Eqs.~\eqref{eq:Af} into Eq.~\eqref{eq:acpdef}, in the limit of $r_f\ll 1$, 
yields
\begin{align} \label{eq:taylor}
    A_{\text{CP}}(f)=& -2\,r_f\,\sin\delta_f\,\sin\phi_f\,+\,\mathcal{O}(r_f^2)~,
\end{align}
requiring both strong ($\delta_f$) and weak ($\phi_f$) relative phases for a non-vanishing direct CP--asymmetry.
Beyond the SM the SCS $D^0$ decay amplitude can be written as
\begin{align}
    \mathcal{A}_f\,=\,\sum_{q=d,s,b}\lambda_q\,(\mathcal{A}^{q}_f)_{\text{SM}}\,+\,\mathcal{A}^{\text{NP}}_f~,\label{eq:amplitude}
\end{align}
where the first term corresponds to the SM contribution with CKM--factors $\lambda_q=V_{cq}^*V_{uq}$ made explicit, and the second term accounts for NP.
Using CKM unitarity $\lambda_d\,+\,\lambda_s\,+\lambda_b\,=\,0$ and
writing for the final states $K^+ K^-$ and $\pi^+ \pi^-$ in the subscripts $f=K$ and $f=\pi$, respectively,
one finds
\begin{align}
\begin{split}\label{eq:Aexpanded}
    \mathcal{A}_{K(\pi)}&=\lambda_{s(d)}\,\big(\mathcal{A}_{K(\pi)}^{s(d)}-\mathcal{A}_{K(\pi)}^{d(s)}\big)_{\text{SM}} \\
    &+\lambda_b\,\big(\mathcal{A}_{K(\pi)}^b-\mathcal{A}_{K(\pi)}^{d(s)}\big)_{\text{SM}}+\mathcal{A}_{K(\pi)}^{\text{NP}}~ .
    \end{split}
\end{align}
Here, the first term is the SCS contribution and the second one corresponds to ``penguin'' contributions with small Wilson coefficients which are strongly CKM--suppressed with respect to the SCS one by $\lambda_{b}/\lambda_{s,d}$. The last term $\mathcal{A}_{K(\pi)}^{\text{NP}}$ encodes NP contributions.
Using Eqs.~\eqref{eq:Af}, \eqref{eq:taylor} and \eqref{eq:Aexpanded}, we obtain
\begin{align}
    \Delta A_{\text{CP}}=\Delta A_{\text{CP}}^{\text{SM}}-\frac{2}{|\lambda_{s,d}|}\,\Delta r^{\text{NP}}~,
\end{align}
where~\footnote{The plus sign between the pion and kaon amplitudes in Eq.~\eqref{eq:rampl} comes from $\lambda_d=-\lambda_s+{\cal{O}}(\lambda_b)$.} 
\begin{align}\label{eq:rampl}
 \begin{split}
    \Delta r^{\text{NP}}&= r_K\sin\delta_K\,\sin\phi_K+r_\pi\sin\delta_\pi\,\sin\phi_\pi~,
\end{split}
\end{align} 
and 
\begin{align}
     r_K=\frac{\mathcal{A}_K^{\text{NP}}}{(\mathcal{A}_K^s-\mathcal{A}_K^d)_{\text{SM}}}~,\quad r_\pi=\frac{\mathcal{A}_\pi^{\text{NP}}}{(\mathcal{A}_\pi^d-\mathcal{A}_\pi^s)_{\text{SM}}}~,
\end{align}
and $r_{\pi,K} \ll 1$.
The strong phases $\delta_{\pi,K}$ are associated with the NP amplitudes.
Since we are interested in maximal NP contributions, we employ in our numerical analysis  $\sin\delta_{\pi,\,K} \sim 1$. Note, there is a priori no information on the sign of $\Delta r^{\text{NP}}$ as it depends on products of  strong and weak phases.
The branching ratios of the $D \to f$ modes are dominated by their respective SM contributions. 
We can therefore extract $\big|(\mathcal{A}_{K\,(\pi)}^{s\,(d)}-\mathcal{A}_{K\,(\pi)}^{d\,(s)})_{\text{SM}}\big|$ from data, see Appendix~\ref{app:fit} for details.

\subsection{CP--violation in \texorpdfstring{$\boldsymbol{D^0}$--$\boldsymbol{\overline{D}^0}$}{D-Dbar} mixing   \label{sec:mixing}}

Here we consider constraints from charm meson mixing.
The  $D^0$--$\overline{D}^0$ transition amplitude can be written as
\begin{align}
    \langle D^0|\mathcal{H}_{\text{eff}}^{\Delta c=2}|\overline{D}^0\rangle\,=\,M_{12}\,-\,\frac{\text{i}}{2}\,\Gamma_{12}~,
\end{align}
which can be parametrized in terms of the following physical quantities
\begin{align}
    x_{12}=2\,\frac{|M_{12}|}{\Gamma}~,~y_{12}=\frac{|\Gamma_{12}|}{\Gamma}~,~ \phi_{12}=\text{arg}\bigg(\frac{M_{12}}{\Gamma_{12}}\bigg)~.
\end{align}
Here, $x_{12}$ and $y_{12}$ are CP--conserving, while $\phi_{12}$ is a phase difference that results in CP--violation in mixing.
A global fit from the HFLAV collaboration \cite{Amhis:2019ckw} results in 
\begin{align}
\begin{split}
    x_{12}&\in[0.22\,,\,0.63]\,\%~,\\
    y_{12}&\in[0.50\,,\,0.75]\,\%~,\\
    \phi_{12}&\in[-2.5^\circ,\,1.8^\circ]~.
    \end{split}\label{eq:Dmixing}
\end{align}
In absence of a sufficiently controlled SM prediction of the mixing parameters, 
we require the NP contributions to saturate the current world averages \eqref{eq:Dmixing}, 
\begin{align}
    x_{12}^{\text{NP}}\leq x_{12}~,\quad  x_{12}^{\text{NP}}\sin\phi_{12}^{\text{NP}}\leq x_{12}\sin\phi_{12}~.\label{eq:Dmixingbounds}
\end{align}

\subsection{CP--violation in \texorpdfstring{$\boldsymbol{c\to u\,\ell^+\ell^-}$}{c to ull}   \label{sec:sl}}

CP--violation in semileptonic rare charm decays arises from complex-valued Wilson coefficients $C^{\ell\ell}_i$, $C^{\ell\ell \, \prime}_i$
in the effective Hamiltonian~\cite{Bause:2019vpr},
\begin{align}
\mathcal{H}_{\text{eff}} \supset -\frac{4\,G_F}{\sqrt2} \frac{\alpha_e}{4\pi}  \sum_{i=9,10} \bigl( C^{\ell\ell}_i O^{\ell\ell}_i + C_i^{\ell\ell\,\prime} O_i^{\ell\ell\,\prime} \bigr) +\,\text{h.c.}\,,
\label{eq:Heffsl}
\end{align}
with the operators
\begin{align}
O_9^{\ell\ell\,(\prime)} &= (\ubar_{L\,(R)} \gamma_\mu c_{L\,(R)}) (\lbar \gamma^\mu \ell) \,, \\ 
O_{10}^{\ell\ell\,(\prime)} &= (\ubar_{L\,(R)} \gamma_\mu c_{L\,(R)}) (\lbar \gamma^\mu \gamma_5 \ell) \,.
\end{align}
Here, $\alpha_e$ denotes the fine structure constant, $G_F$ is Fermi's constant and $L=(1-\gamma_5)/2, R=(1+\gamma_5)/2$ are chiral projectors.
CP--violation has not been observed in semileptonic $\vert \Delta c \vert=\vert \Delta u\vert=1$ decays yet. Available measurements for CP--asymmetries in rare semileptonic charm decays are at the level of few to $\mathcal{O}(10)\,\%$~\cite{Aaij:2018fpa}, which is close to possible NP effects~\cite{Fajfer:2012nr,deBoer:2018buv,Bause:2019vpr}.

Branching ratio and high--$p_T$ data imply the following constraints, barring cancellations~\cite{deBoer:2015boa,Fuentes-Martin:2020lea}
\begin{align} \label{eq:limits}
\vert   C^{\mu \mu \, ( \prime)}_{9,10} \vert \lesssim 1  \, , \quad \vert   C^{ee \, ( \prime)}_{9,10} \vert \lesssim 3 \, ,
\end{align}
stronger for muons than for electrons. 

\section{A flavorful \texorpdfstring{$\boldsymbol{Z^\prime}$}{Z'} in charm}\label{sec:Zp}

We work out NP--effects in charm 
from anomaly-free  $U(1)^\prime$--extensions of the SM with fermion charges $F_{\psi_i}$ that depend on the generation, $i=1,2,3$.
Specifically, SM fermion multiplets plus possibly right-handed neutrinos $\psi=Q,u,d,L,e,\nu$ in representations of $SU(3)_C \times SU(2)_L \times U(1)_Y \times U(1)^\prime$ can be 
characterized, in that order, as 
\begin{align} 
\begin{split}
Q_i & = (3, 2, 1/6,F_{Q_i})~,  \quad
u_i = (3, 1, 2/3, F_{u_i})~,  \quad \\
d_i &= (3, 1, -1/3, F_{d_i})~,~ 
L_i  = (1, 2, -1/2, F_{L_i})~,   \quad \\
e_i & =(1, 1, -1,F_{e_i})~,~~  \quad 
\nu_i = (1, 1, 0,F_{\nu_i})~.
\end{split}
\end{align}
Concrete models with $F_{\psi_i}$--assignments  that fulfill the anomaly-cancellation conditions and induce $c \to u$ flavor changing neutral currents (FCNCs) are given in TABLE~\ref{tab:ZprimeModels}. 
Related models (models 1 to 8) have been studied previously  in the context of semileptonic rare charm decays in Ref.~\cite{Bause:2019vpr}, to which we refer for further details.
The models in TABLE~\ref{tab:ZprimeModels} satisfy 
$ \sum_{i=1}^{3}\left( F_{Q_i} -F_{L_i} +2\,F_{u_i} -F_{d_i} -F_{e_i} \right) = 0$ and therefore avoid kinetic mixing at one-loop~\cite{Holdom:1985ag}.

\begin{table*}[ht!]
	\centering
	\resizebox{0.8\textwidth}{!}{%
	\begin{tabular}{c|rcr|rcr|rcr|rcr|rcr|rcr}
	\hline \hline
		model &&$F_{Q_i}$&&& $F_{u_i}$ &&& $F_{d_i}$ &&& $F_{L_i}$ &&& $F_{e_i}$ &&& $F_{\nu_i}$ &\\
		\hline
        2 & \phantom{-}3 & \phantom{-}3 & -6 & -8 & \phantom{-}4 & 4 & -10 & 10 & 0 & -6 & 5 & 1 & 0 & \phantom{-}0 & 0 & 0 & \phantom{-}0 & \phantom{-}0 \\
        4 & -1 & -1 & \phantom{-}2 & -1 & \phantom{-}2 & -1 & 0 & \phantom{-}0 & 0 & -1 & 1 & 0 & -2 & \phantom{-}2 & 0 & -2 & -1 & 3\\
        5 & -1 & -1 & \phantom{-}2 & -1 & \phantom{-}2 & -1 & 2 & -1 & -1 & -1 & 1 & 0 & -1 & \phantom{-}1 & 0 & 0 & \phantom{-}0 & 0\\
        9 &  \phantom{-}0 & \phantom{-}0  & \phantom{-}0 & -11 & -2 & 13 & 7 & \phantom{-}7 & -14 & -8 & 3 & 5 & -6 & 16 & -10 & 0 & \phantom{-}0 & 0\\
        10 & \phantom{-}0 & \phantom{-}0  & \phantom{-}0 & -13 & \phantom{-}6 & 7 & -1 & -14 & 15  & -15 & 15 & 0 & -14 & 18 & -4 & 0 & \phantom{-}0 & 0\\
        10$\mu$ & \phantom{-}0 & \phantom{-}0  & \phantom{-}0 & -13 & \phantom{-}6 & 7 & -1 & -14 & 15  & -15 & 0 & 15 & -14 & -4 & 18 & 0 & \phantom{-}0 & 0\\
		\hline \hline
	\end{tabular}
	}	
		\caption{Sample solutions of an anomaly-free $U(1)^\prime$--extension of the SM$+3\,\nu_R$ with $F_{Q_1}=F_{Q_2}$. Models 2, 4 and 5 are taken from Ref.~\cite{Bause:2019vpr}. Models 9 and 10 feature $F_{Q_i}=0$. In general, the ordering of generations is arbitrary due to permutation invariance. However, our analysis explicitly uses the ordering stated here, that is, the 
		$i$th entry corresponds to the  $i$th generation. Model 10$\mu$ is the same as model 10 with the smallest lepton-coupling to muons.}
		\label{tab:ZprimeModels}
\end{table*}

In Section~\ref{sec:g} we discuss couplings  of the fermions to the $Z^\prime$--boson, which arises from the  $U(1)^\prime$--group.
We assume the $Z^\prime$ to have a mass $M_{Z^\prime}$ of the electroweak scale or heavier.
We discuss the induced $c \to u$ four-quark operators and Wilson coefficients in  Section~\ref{sec:ops}.
In Section~\ref{sec:DDbar} we discuss how to bypass constraints from $D^0$--$\overline{D}^0$ mixing.
We work out predictions for $\Delta A_{\text{CP}}$ in Section~\ref{sec:DACP}.

\subsection{\texorpdfstring{$\boldsymbol{Z^\prime}$}{Z'}--FCNCs  \label{sec:g}}

The $Z^\prime$--couplings relevant to charm FCNCs can be written as
\begin{align}
\begin{split}
    \mathcal{L}_{Z^\prime} \supset
    &\left( g_L^{uc}\,\bar u_L \gamma^\mu c_L Z^\prime_\mu + g_R^{uc}\,\bar u_R \gamma^\mu c_R Z^\prime_\mu + \text{h.c.} \right)\\
    &+\phantom{(} g^d_L\,\bar d_L \gamma^\mu d_L Z^\prime_\mu + g^d_R\,\bar d_R \gamma^\mu d_R Z^\prime_\mu  \\
    &+\phantom{(}g^s_L\,\bar s_L \gamma^\mu s_L Z^\prime_\mu + g^s_R\,\bar s_R \gamma^\mu s_R Z^\prime_\mu\\
    &+\phantom{(}g_L^{\ell\ell}\bar \ell_L \gamma^\mu \ell_L Z^\prime_\mu+g_R^{\ell\ell}\bar \ell_R \gamma^\mu \ell_R Z^\prime_\mu\,,
\end{split}    \label{eq:LintZ}
\end{align}
with $\ell=e,\,\mu,\,\tau$. 
The flavor diagonal couplings $g^{d,s}_{L,R}$ and $g^{\ell\ell}_{L,R}$ are given as the $U(1)^\prime$--gauge coupling $g_4$ times the associated charge $F_\psi$.

The $|\Delta c|=|\Delta u|=1$ FCNC couplings $g_{L,R}^{uc}$ are generated via rotations from the gauge to the mass basis, and are in general complex-valued.
Four different unitary rotations exist in the quark sector, corresponding to the left-handed (LH) and right-handed (RH) ones both for up- and down-type quarks. 
The product of LH up- and down-type rotations gives the CKM--matrix. In order to evade the severe constraints in the kaon sector, we assume the CKM--matrix to predominantly stem from the LH up-type rotation, implying
\begin{align}
    g^{uc}_L\approx g_4\,\lambda_{\text{CKM}}\,\Delta F_L~,\, \quad   \Delta F_L =F_{Q_2}- F_{Q_1}~,\,\label{eq:gucleft}
\end{align}
where $\lambda_{\text{CKM}}\sim 0.2$ denotes the Wolfenstein parameter and we used $\lambda_{\text{CKM}} \ll 1$.
In contrast, the RH rotation is a priori unconstrained and induces
\begin{align} 
g_R^{uc}=g_4 \sin \theta_u \cos \theta_u\,e^{\text{i} \phi_R} \Delta F_R \, , 
\label{eq:gucright}
\end{align}
where $\theta_u$ is the up-charm mixing angle for the up-quark singlets, $\Delta F_R =F_{u_2}- F_{u_1}$ and $\phi_R$ the corresponding CP--phase.

\subsection{Four-fermion operators and matching \label{sec:ops}}

Generation-dependent quark-couplings result in additional operators in the effective weak Hamiltonian beyond the ones considered usually, \textit{i.e.}~Ref.~\cite{Altmannshofer:2012ur}. At the scale $m_b<\mu<\mu_{\text{EWK}}$~, 
\begin{align}
    \mathcal{H}_{\text{eff}}^{|\Delta \text{c}|=1}\:\supset\:\frac{G_F}{\sqrt{2}}\,\sum_i\,\widetilde{C}_i^{(\prime)}\,\widetilde{Q}_i^{(\prime)}\,+\,\text{h.c.}~,\label{eq:DC1Hamiltonian}
\end{align}
with the new operators
\begin{align}
\widetilde{Q}_7 &= (\bar u c)_{V-A} \sum_{q}F_{u_i,d_i}\, (\bar q q)_{V+A}~,\label{eq:OpZ7}\\
\widetilde{Q}_7^\prime &= (\bar u c)_{V+A} \sum_{q} F_{Q_i}\, (\bar q q)_{V-A}~, \\
\widetilde{Q}_8 &= (\bar u_\alpha c_\beta)_{V-A} \sum_{q}F_{u_i,d_i}\, (\bar q_\beta q_\alpha)_{V+A}~, \\
\widetilde{Q}_8^{\prime} &= (\bar u_\alpha c_\beta)_{V+A} \sum_{q}F_{Q_i}\, (\bar q_\beta q_\alpha)_{V-A}~, \\
\widetilde{Q}_9 &= (\bar u c)_{V-A} \sum_{q}F_{Q_i}\, (\bar q q)_{V-A}~, \\
\widetilde{Q}_9^\prime &= (\bar u c)_{V+A} \sum_{q} F_{u_i,d_i}\, (\bar q q)_{V+A}~,\\
\widetilde{Q}_{10} &= (\bar u_\alpha c_\beta)_{V-A} \sum_{q}F_{Q_i}\, (\bar q_\beta q_\alpha)_{V-A}~,  \\
    \widetilde{Q}_{10}^{\prime} &= (\bar u_\alpha c_\beta)_{V+A} \sum_{q}F_{u_i,d_i}\, (\bar q_\beta q_\alpha)_{V+A}~,\label{eq:OpZ10}
\end{align}
where $(V\pm A)$ refers to the Dirac structures
$\gamma_\mu (1 \pm \gamma_5)$, $q=u,c,d,s,b$ and $\alpha,\beta$ are the color indices.
The strength of these operators is given by their respective Wilson coefficients $\widetilde{C}_i$, $\widetilde{C}_i^\prime$ which depend on both heavy masses and weak phases responsible for CP--violating phenomena. The Wilson coefficients induced by the Lagrangian~\eqref{eq:LintZ} read
\begin{align}
\begin{split}\label{eq:WCZp}
    \widetilde{C}_7\left(M_{Z^\prime}\right) &= \widetilde{C}_9\left(M_{Z^\prime}\right)= \frac{\sqrt{2}}{G_F}\,g_L^{uc}\,\frac{g_4}{4\,M_{Z^\prime}^2} \,, \\
    \widetilde{C}_7^\prime\left(M_{Z^\prime}\right) &= \widetilde{C}_9^\prime\left(M_{Z^\prime}\right) = \frac{\sqrt{2}}{G_F}\,g_R^{uc}\,\frac{g_4}{4\,M_{Z^\prime}^2} \, ,\\
    \widetilde{C}_8^{(\prime)} \left(M_{Z^\prime}\right)&  =\widetilde{C}_{10}^{(\prime)} \left(M_{Z^\prime}\right)=0 \, .
\end{split}
\end{align}
They are evolved from $M_{Z^\prime}$ to $m_c$ using the renormalization group equations (RGEs) with top and bottom quarks integrated out at their respective threshold scales. Finite values of $\widetilde{C}_8^{(\prime)}$ and $\widetilde{C}_{10}^{(\prime)}$ arise from the RGE mixing at the charm mass scale, see Appendix~\ref{app:A} for details.

\subsection{\texorpdfstring{$\boldsymbol{D^0-\overline{D}^0}$}{D-Dbar} mixing constraints}\label{sec:DDbar}

Rare  $\vert \Delta c \vert=\vert \Delta u\vert=1$ decays are induced in the $Z^\prime$--models by operators with coefficients proportional to $g_L^{uc}$ or $g_R^{uc}$ in Eq.~\eqref{eq:WCZp}.
These couplings induce at second order  $D^0$--$\overline{D}^0$ mixing \eqref{eq:Dmixing}, 
and are constrained as 
\begin{align}
\label{eq:quadraticform}
\begin{split}
 &{\left\vert (g_L^{uc})^2+(g_R^{uc})^2-X\,{g_L^{uc}}\,{g_R^{uc}} \right\vert} \lesssim 6\cdot 10^{-7}  \left(\frac{M_{Z^\prime}}{\TeV}\right)^2, 
\end{split}
\end{align}
with $X\sim20$ for $M_{Z^\prime}$ in the TeV range \cite{Bause:2019vpr}.
This constraint on $x_{12}$ can be evaded  if both $g_L^{uc}$ and $g_R^{uc}$ are present, for
either  $g_L^{uc} \sim X g_R^{uc}$   or $g_L^{uc} \sim 1/X g_R^{uc}$.
However, in these cases the CP--phases have to be aligned $\text{Arg}(g_L^{uc}) \sim \text{Arg} (g_R^{uc})$ to fulfill Eq.~\eqref{eq:quadraticform}. As kaon constraints force 
$\text{Arg}(g_L^{uc})$ to be SM--like,  CP--violating effects in charm become negligible. 

We therefore choose $g_L^{uc}=0$, which can be achieved with $\Delta F_L=0$.
The models in TABLE~\ref{tab:ZprimeModels} satisfy for this reason $F_{Q_1}=F_{Q_2}$.
Consequently, we focus on FCNCs in the up-singlet sector \eqref{eq:gucright}, that is, $g_R^{uc} \neq 0$ and complex.

\begin{figure}[!t]
    \centering 
\includegraphics[width=0.45\textwidth]{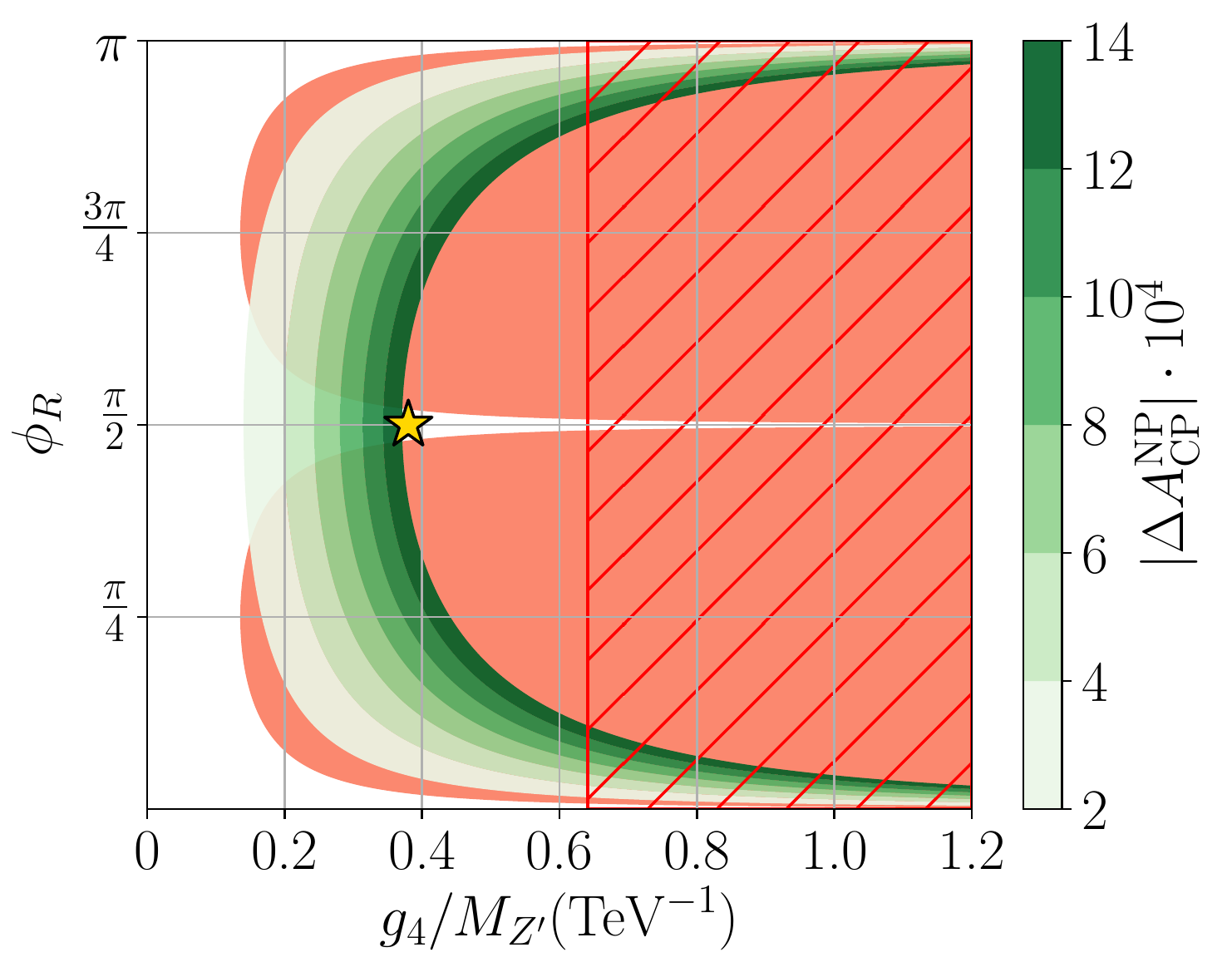}
\caption{$|\Delta A_{\text{CP}}^{\text{NP}}|$ (green bands) versus $D^0$--$\overline{D}^0$ mixing exclusion regions \eqref{eq:Dmixingbounds} on the imaginary part $x_{12} \sin  \phi_{12}$ (red area)   and the absolute value $x_{12}$ (red hatched area) in the $\phi_R$--$g_4/M_{Z^\prime}\,(\TeV^{-1})$ plane for $\theta_u=1\cdot10^{-4}$.
$F_\psi$--charges are as in model 2, see TABLE~\ref{tab:ZprimeModels}. The golden star indicates a benchmark point \eqref{eq:bench}, see text for details.}
\label{fig:imagesDeltaACPweakphase}
\end{figure}

If there is a single coupling only, the above mixing constraint on $x_{12}$ becomes
\begin{align}\label{eq:g-bound}
|g_A^{uc}| \lesssim 8 \cdot 10^{-4}  \left(\frac{M_{Z^\prime}}{\TeV}\right) \,, \quad A=L,R ~.
\end{align}
The even tighter constraint \eqref{eq:Dmixingbounds} for CP--violating couplings on $x_{12} \sin  \phi_{12}$ can be bypassed for $\text{Arg} (g_R^{uc})=\phi_R$
around $\pi/2$ (or $3 \pi/2$), as the CP--phase of the mixing amplitude is twice the one of the $|\Delta c|=|\Delta u|=1$ FCNC~\cite{Altmannshofer:2012ur}.
The contributions to $\Delta A_{\text{CP}}$ become maximal while simultaneously mixing constraints are satisfied.  
This interplay of $\phi_R$ versus the coupling $g_4/M_{Z^\prime}\,(\TeV^{-1})$ for model 2 and fixed  $\theta_u=1\cdot10^{-4}$ is illustrated in FIG.~\ref{fig:imagesDeltaACPweakphase}.
The red (hatched) area corresponds to the $D^0$--$\overline{D}^0$ mixing constraints on the imaginary part $x_{12} \sin \phi_{12}$ (absolute value $x_{12}$).
$Z^\prime$--induced values of $\Delta A_{\text{CP}}$ are shown in green.
Indeed the region around $\phi_R\sim\pi/2$ is viable and can induce  $\Delta A_{\text{CP}}^{\text{NP}} \sim 10^{-3}$.

\subsection{\texorpdfstring{$\boldsymbol{Z^\prime}$}{Z'}--effects for \texorpdfstring{$\boldsymbol{\Delta A_{\text{CP}}}$}{Delta ACP}}\label{sec:DACP}

Taking into account the running from $M_{Z^\prime}$ to $m_c$, details of which are given in Appendix~\ref{app:A}, we find that $\Delta A_{\text{CP}}$ can be written as
\begin{align}\label{eq:acp_zprime_formula}
\begin{split}
\Delta A_{\text{CP}}^{\text{NP}} &=  A_{\text{CP}}^{\text{NP}}\left(K^+K^-\right) - A_{\text{CP}}^{\text{NP}}\left(\pi^+\pi^-\right)~,
\end{split}
\end{align}
with
\begin{align}
\begin{split}
 A_{\text{CP}}^{\text{NP}}\left(K^+K^-\right) &\sim \frac{g_4^2}{M_{Z^\prime}^2}\theta_u\,\Delta F_R \left[c_K\,F_{Q_2} +d_K\,F_{d_2}\right]~,\\
A_{\text{CP}}^{\text{NP}}\left(\pi^+\pi^-\right) &\sim \frac{g_4^2}{M_{Z^\prime}^2}\theta_u\,\Delta F_R   \left[c_\pi\,F_{Q_1} +d_\pi\,F_{d_1}\right]~,  
\end{split}
\label{eq:acp_zprime}
\end{align}
where
\begin{align}\label{eq:cd_piK}
\begin{split}
    c_K&=\frac{\chi_K}{a_K}\,r_1(m_c,M_{Z^\prime})~,~ c_\pi=-\frac{\chi_\pi}{a_\pi}\,r_1(m_c,M_{Z^\prime})~,\\
    d_K&=\frac{1}{a_K}\,r_2(m_c,M_{Z^\prime})~,~
    d_\pi=-\frac{1}{a_\pi}\,r_2(m_c,M_{Z^\prime})~.  
\end{split}
\end{align}
As explained in the previous Section \ref{sec:DDbar}, we analyze models with $g^{uc}_L=0$ and  Im$(g^{uc}_R)$ large.
In Eq.~\eqref{eq:acp_zprime} we use $\sin \delta_{\pi,K}\,\sin \phi_R \sim 1$ and anticipated $\theta_u \ll 1$. The parameters $c_{K,\pi}$ and $d_{K,\pi}$ depend on the chiral factors $\chi_{K,\pi}$ at the charm scale, the LO QCD running functions $r_{1,2}(m_c,M_{Z^\prime})$  and the tree-level contributions $a_{K,\pi}$, which are determined experimentally.
Further details can be found in Appendices~\ref{app:fit}--\ref{app:RGE}. Numerical values of $c_{K,\pi}$ and $d_{K,\pi}$ for different $Z^\prime$ masses are displayed in TABLE~\ref{tab:constants_acp}.
\begin{table}[ht!]
\renewcommand*{\arraystretch}{1.2}
 \centering
 \begin{tabular}{cccccccccccc}
  \hline
  \hline
 $M_{Z^\prime}$ [TeV] && $2$ && $4$ && $6$ && $8$ && $10$ \\
  \hline
 $c_K$ && $\phantom{-}1.133$ && $\phantom{-}1.217$ &&  $\phantom{-}1.266$ &&  $\phantom{-}1.302$ &&$\phantom{-}1.330$\\
 $d_K$ && $-0.046$ && $-0.054$ && $-0.058$ && $-0.061$ && $-0.063$\\
 $c_\pi$ && $-1.446$ && $-1.553$ && $-1.616$ && $-1.661$ && $-1.698$ \\
 $d_\pi$ && $\phantom{-}0.058$ && $\phantom{-}0.068$ && $\phantom{-}0.074$ && $\phantom{-}0.077$ && $\phantom{-}0.080$ \\
  $d_{\pi^\prime}$ && $\phantom{-}0.071$ && $\phantom{-}0.083$ && $\phantom{-}0.090$ && $\phantom{-}0.094$ && $\phantom{-}0.098$ \\
$d_{\pi^0}$ && $\phantom{-}0.077$ && $\phantom{-}0.090$ && $\phantom{-}0.097$ && $\phantom{-}0.102$ && $\phantom{-}0.106$ \\
  \hline
  \hline
  \end{tabular}
   \caption{Parameters $c_{K,\pi}$,\,$d_{K,\pi}$ and $d_{\pi^\prime,\pi^0}$ in $({\rm TeV})^2$ as defined in Eq.~\eqref{eq:cd_piK} and Eq.~\eqref{eq:dpiprimedef}, respectively, for different $Z^\prime$ masses.}
   \label{tab:constants_acp}
\end{table}
\begin{figure*}[!t]
    \centering 
\includegraphics[width=0.48\textwidth]{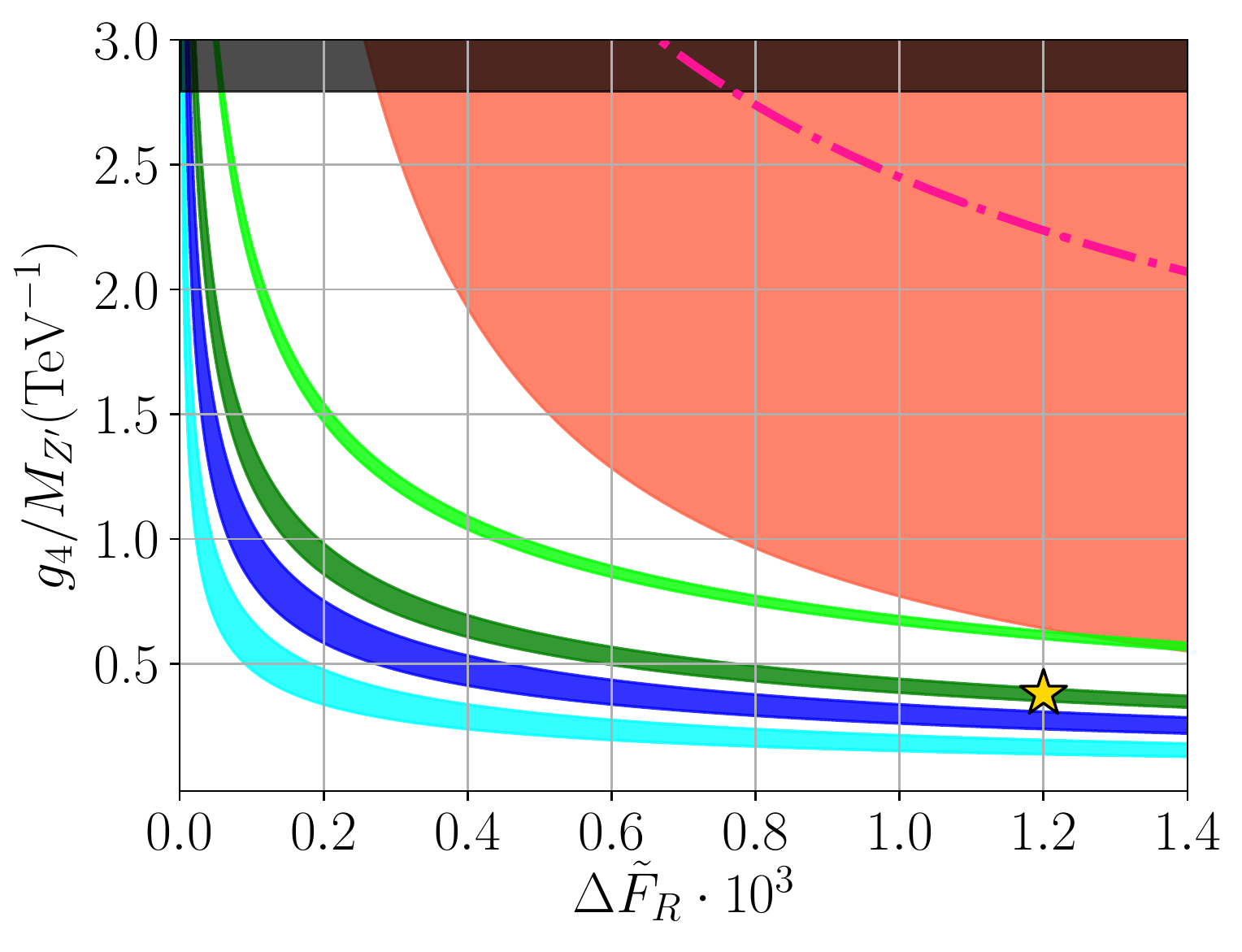}
\includegraphics[width=0.48\textwidth]{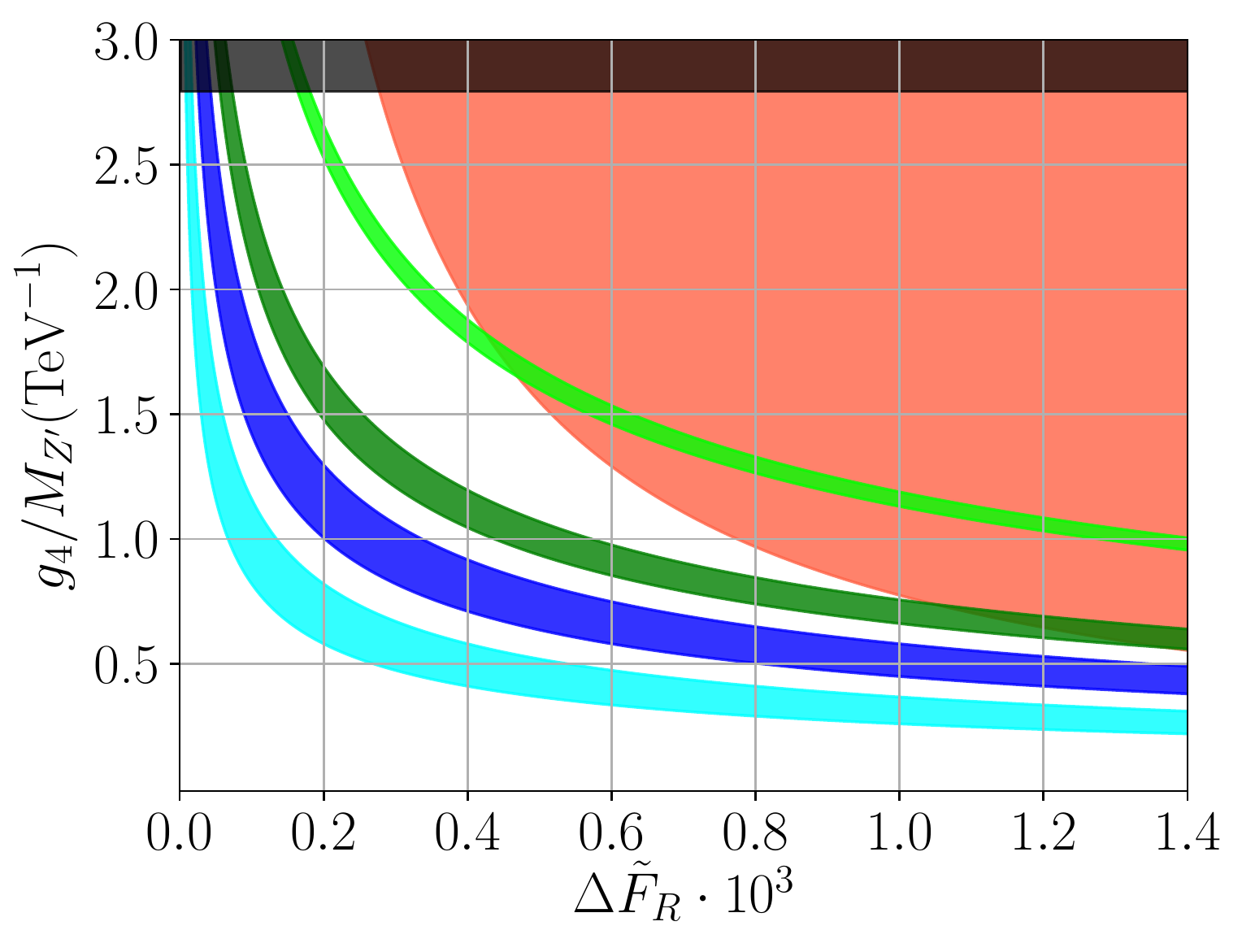}
\includegraphics[width=0.48\textwidth]{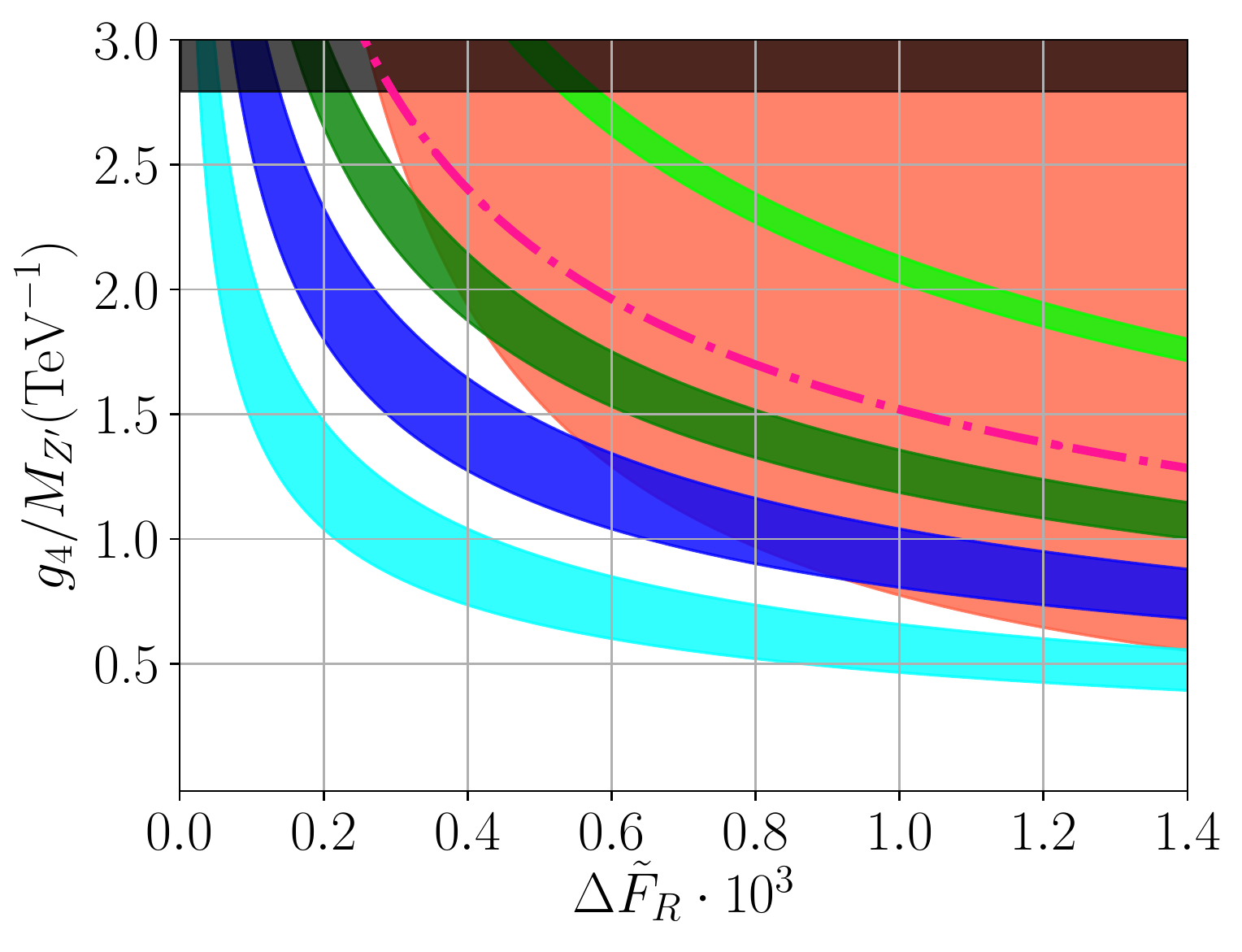}
\includegraphics[width=0.48\textwidth]{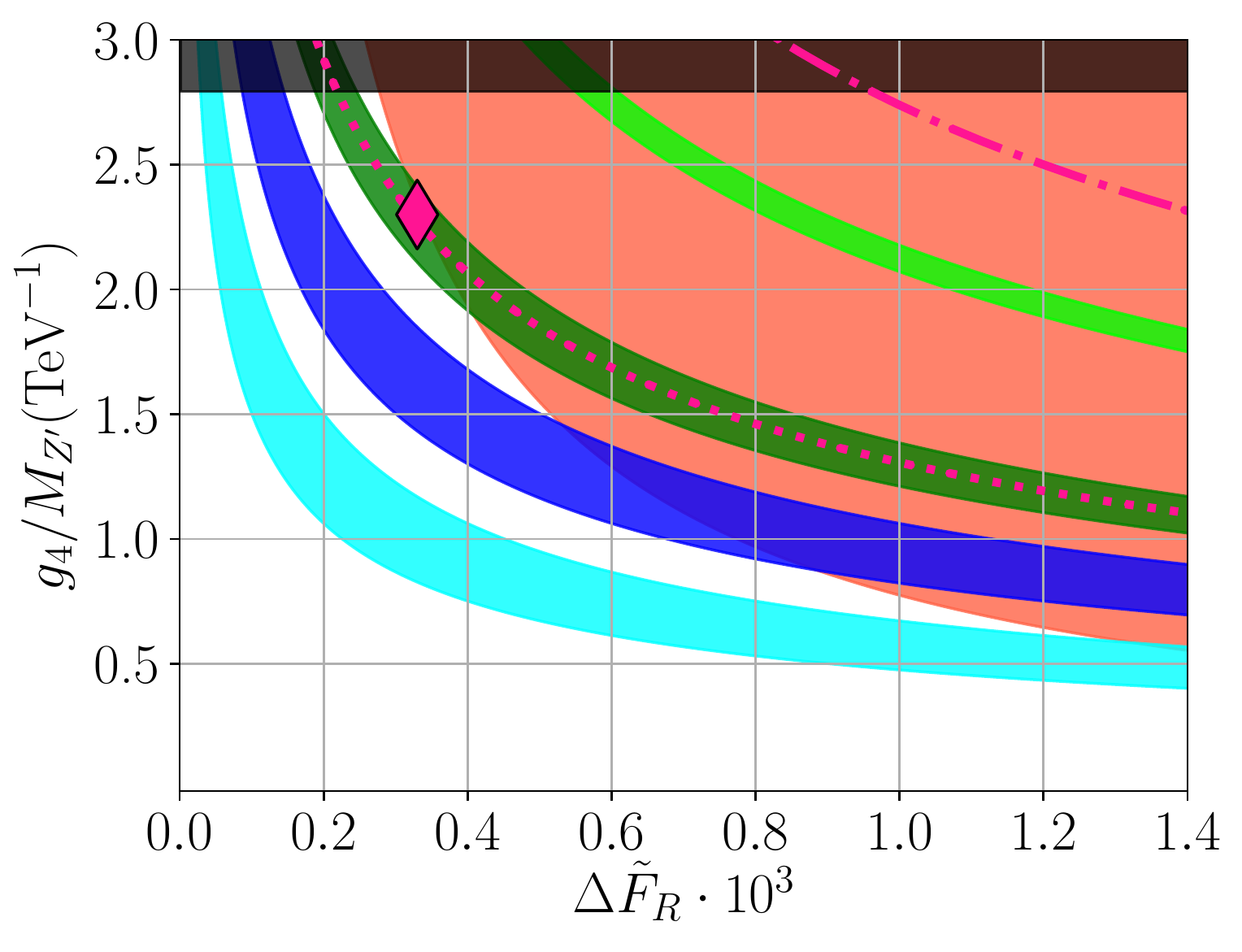}
\caption{$\vert\Delta A_{\text{CP}}^{\text{NP}}\vert$ for different $Z^\prime$--models (2 upper left, 5 upper right, 9 lower left and 10($\mu$) lower right) in the plane of  $g_4/M_{Z^\prime}\,(\TeV^{-1})$ and $\Delta\widetilde{F}_R=\Delta F_R\cdot \theta_u$, together with the excluded region from $D^0$--$\overline{D}^0$ mixing (red). Light green, dark green, blue and cyan bands correspond to ${\vert\Delta A_{\text{CP}}^{\text{NP}}\vert}=(4.0\pm0.2)\cdot10^{-3}$,  ${\vert\Delta A_{\text{CP}}^{\text{NP}}\vert}=(1.5\pm0.2)\cdot10^{-3}$, ${\vert\Delta A_{\text{CP}}^{\text{NP}}\vert}= (8\pm2)\cdot10^{-4}$ and ${\vert\Delta A_{\text{CP}}^{\text{NP}}\vert}=(3\pm1)\cdot10^{-4}$, respectively. The black region indicates the upper bound coming from perturbativity and direct searches in dimuon and dielectron spectra~\cite{Tanabashi:2018oca}, which read $g_4\leq4\pi\,$ and $M_{Z^\prime} \geq4.5\,\TeV$, respectively. The magenta dash--dotted and dotted lines show  the stronger (if any) of the bounds from Eqs.~\eqref{eq:bounds_brsemimuonic} and \eqref{eq:bounds_brmuonic}. In the lower right plot the dotted  line corresponds to model 10, and the dash--dotted to model 10$\mu$. The golden star and pink diamond are benchmark points \eqref{eq:bench} and 
\eqref{eq:bench10}. See text for details.}
\label{fig:imagesDeltaACP}
\end{figure*}

In FIG.~\ref{fig:imagesDeltaACP} we show sizable $Z^\prime$--contributions to $\Delta A_{\text{CP}}^{\text{NP}}$  and  $D^0$--$\overline{D}^0$ mixing constraints (red area) 
in the plane of 
$g_4/M_{Z^\prime}\,(\TeV^{-1})$ and the parameter $\Delta \widetilde{F}_R=\Delta F_R\,\theta_u$ for models 2, 5, 9 and 10($\mu$).
The corresponding plot of model 4 is not given in FIG.~\ref{fig:imagesDeltaACP}  because it exhibits very similar  bands as model 5 due to  identical $F_{Q_{1,2}}$ and $\Delta F_R$,
as shown in TABLE~\ref{tab:ZprimeModels}. 
Constraints from branching ratios of (semi-)muonic $D$--decays (dash-dotted and dotted lines), here for $g_L^{uc}=0$, \cite{Bause:2019vpr}
\begin{align}
\begin{split}\label{eq:bounds_brsemimuonic}
 \vert g_R^{uc}&\vert  \sqrt{(g_L^{\mu\mu})^2 + (g_R^{\mu\mu})^2} \\  &=g_4^2 \vert \Delta \widetilde{F}_R\vert \sqrt{F^2_{L_2}+ F^2_{e_2}} \lesssim 0.04 \left({M_{Z^\prime} \over {\rm TeV}}\right)^2 ,
 \end{split}\\
 \begin{split}\label{eq:bounds_brmuonic}
 \bigl|g_R^{uc}& (g_L^{\mu\mu} - g_R^{\mu\mu}) \big| \\ &=g_4^2 | \Delta \widetilde{F}_R (F_{L_2}- F_{e_2})| \lesssim 0.03 \left({M_{Z^\prime} \over {\rm TeV}}\right)^2 ,
 \end{split}
\end{align}
start to be competitive with mixing constraints close to the non-perturbativity region (black region).  
This is particularly relevant for model 9 and 10, which
exhibit large couplings to leptons. 
To evade the muon constraints and allow for slightly larger values of $\Delta A_{\rm CP}$ we also consider model 10$\mu$, which is the same as model 10 with the lepton-charges ordered in such a way that the smallest ones are for muons, stressing the interplay between hadronic and leptonic sectors; model 10 can accommodate  $\Delta A_{\text{CP}}^{\text{NP}}$ up to {$1.5\cdot 10^{-3}$}, while model 
10$\mu$ can reach {$1.8\cdot 10^{-3}$}.  FIG.~\ref{fig:imagesDeltaACP} shows the stronger bound for each model, \textit{i.e.},~Eq.~\eqref{eq:bounds_brmuonic} for models 2, 5, 9 and 10$\mu$ (dash-dotted) and Eq.~\eqref{eq:bounds_brsemimuonic} for model 10 (dotted).

In FIGS.~\ref{fig:imagesDeltaACPweakphase} and~\ref{fig:imagesDeltaACP} we show benchmark points. They pass constraints from $D$--mixing and  semi(-muonic) decays,
while giving $\Delta A_{\text{CP}}^{\text{NP}} \sim 10^{-3}$.
The golden star corresponds to  model 2 with $\Delta F_R=12$ and
\begin{align} \label{eq:bench}
\phi_R \sim \pi/2 \,, \,\, g_4/M_{Z^\prime} \sim 0.38/\TeV \,, \,\, \theta_u \sim  1\cdot10^{-4}~.
\end{align}
The pink diamond corresponds to model 10$\mu$ with $\Delta F_R=19$ and 
\begin{align} \label{eq:bench10}
\phi_R \sim \pi/2 \,, \,\, g_4/M_{Z^\prime} \sim 2.3/\TeV \,, \,\, \theta_u \sim 1.7\cdot10^{-5}~.
\end{align}

We learn that $Z^\prime$--models with charges as in TABLE~\ref{tab:ZprimeModels} can provide concrete NP--interpretations of $\Delta A_{\text{CP}}$ of the order of $10^{-3}$. $D^0$--$\overline{D}^0$ mixing provides upper limits on the achievable $\Delta A_{\text{CP}}^{\text{NP}}$.
To distinguish the different model scenarios  we explore correlations of $\Delta A_{\text{CP}}$ with other sectors, hadronic 2-body $D$--decays in Section~\ref{sec:patterns}
and semileptonic $c \to u\, \ell^+ \ell^-$ transitions in Section~\ref{sec:slCP}.

\section{Patterns in hadronic decays}\label{sec:patterns}

$Z^\prime$--models with non-universal charges $F_\psi$ can give rise to large flavor-breaking effects which could explicitly violate relations between hadronic charm decays~\cite{Grossman:2013lya,Grossman:2018ptn, Grossman:2019xcj, Muller:2015rna}. We study signatures of $Z^\prime$--induced
U--spin and isospin breaking in Section~\ref{subs:uspin} and Section~\ref{sec:I}, respectively.
$A_{\text{CP}}$ in $D^0 \to \pi^0 \pi^0$ is studied in Section~\ref{sec:pi0}.

\subsection{U--spin patterns in $\texorpdfstring{\boldsymbol{D^0\to \pi^+ \pi^-,\,K^+K^-}}{D to PiPiorKK}$}\label{subs:uspin}

\begin{figure*}[!t]
    \centering
\includegraphics[width=0.45\textwidth]{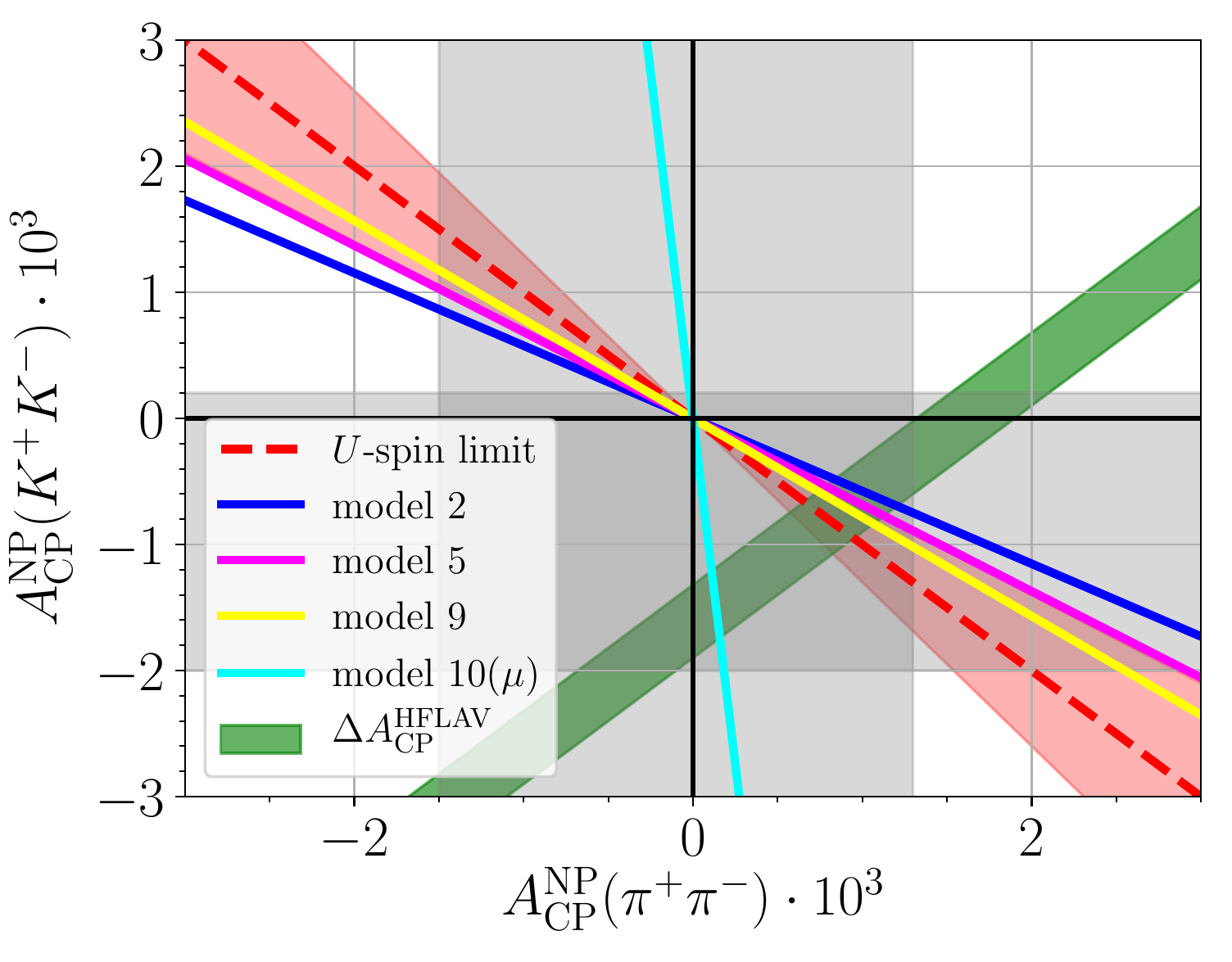}
\caption{$A_{\text{CP}}(K^+K^-)$ versus  $A_{\text{CP}}(\pi^+\pi^-)$ with predictions in the $Z^\prime$--models 2, 5, 9 and 10($\mu$) in blue, magenta, yellow and cyan lines, respectively. 
The green band corresponds to the experimental world average of $\Delta A_{\text{CP}}$ \eqref{eq:worldaverage} at 1 $\sigma$. 
The gray bands indicate the present experimental $1\sigma$ regions given in TABLE~\ref{tab:CP--data}.
The U--spin limit \eqref{eq:UB1} (red dashed line) and $\lesssim 30$ \% SM-like U--spin breaking (red area) is also shown.}
\label{fig:uspin}
\end{figure*}
\begin{figure*}[!t]
    \centering
\includegraphics[width=0.45\textwidth]{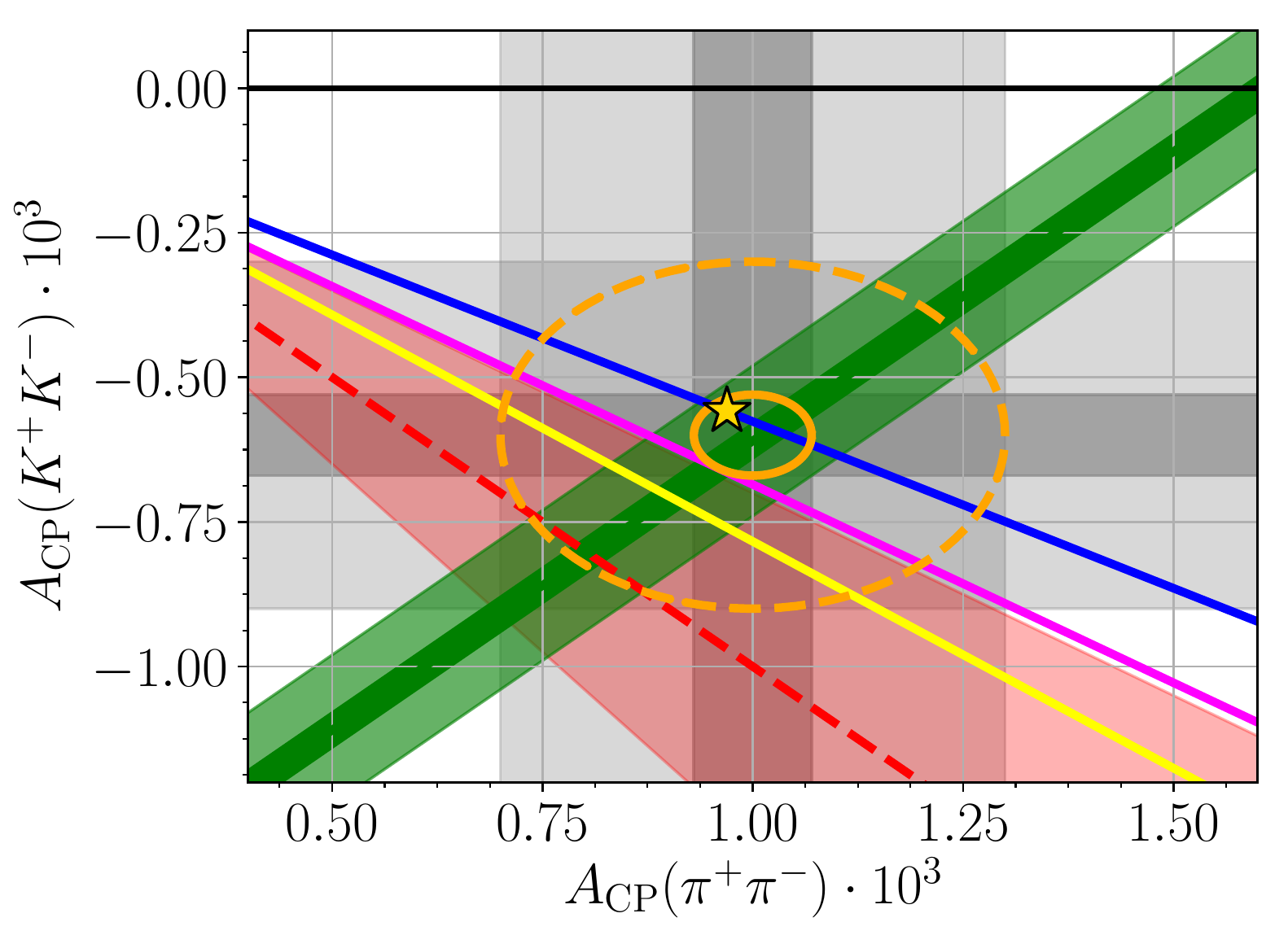}
\includegraphics[width=0.45\textwidth]{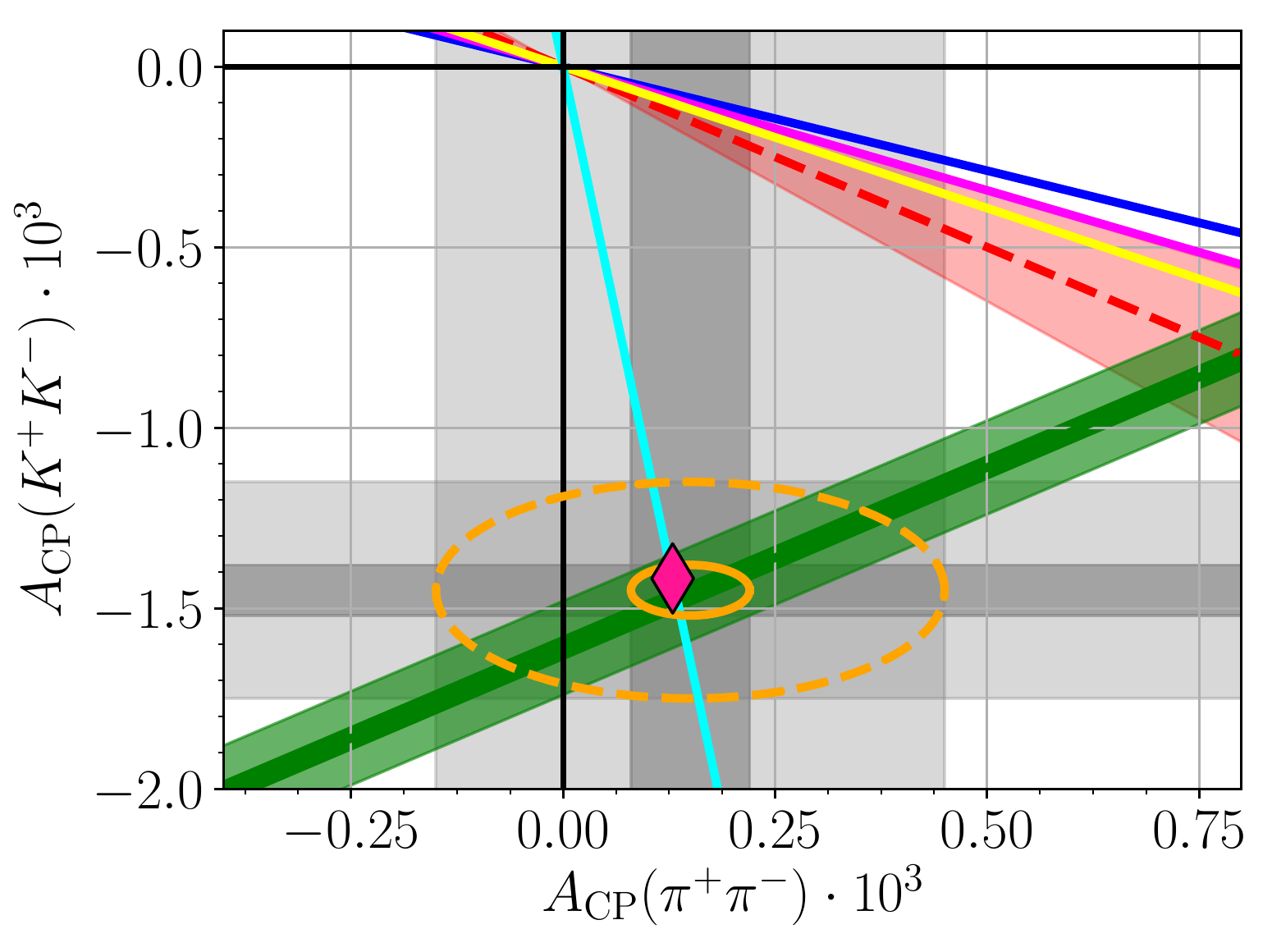}
\caption{Future projections for $A_{\text{CP}}(K^+K^-)$ versus  $A_{\text{CP}}(\pi^+\pi^-)$  with predictions in the $Z^\prime$--models 2, 5, 9 and 10($\mu$) in blue, magenta, yellow and cyan lines,  respectively. The green band corresponds to  the central value of the present experimental world average of $\Delta A_{\text{CP}}$ \eqref{eq:worldaverage} with future 1 $\sigma$ sensitivities according to TABLE~\ref{tab:CP--data}.
The gray bands illustrate two future  measurements of the individual asymmetries. The central values are given in Eq.~\eqref{eq:centralvals}, the uncertainties are scaled according to TABLE~\ref{tab:CP--data}. Lighter (darker) bands correspond to LHCb Run 1-3 (1-5). Assuming for simplicity gaussian errors a dashed (solid) ellipse  occurs around model 2 (plot to the left) and model 10$\mu$ (plot to the right) for LHCb Run 1-3 (1-5). 
The U--spin limit \eqref{eq:UB1} (red dashed line) and $\lesssim 30$ \% SM-like U--spin breaking (red area) is also shown.
The golden star and pink diamond are benchmark points \eqref{eq:bench} and \eqref{eq:bench10}.}
\label{fig:uspinfuture}
\end{figure*}
U--spin breaking arises for $F_{Q_1} \neq F_{Q_2}$ or $F_{d_1} \neq F_{d_2}$, and can upset the U--spin sum rule~\cite{Grossman:2013lya}
\begin{align}
A_{\text{CP}}(D^0  &\to K^+ K^-)+A_{\text{CP}}(D^0 \to \pi^+ \pi^-)=0 \, . \label{eq:UB1} 
\end{align}
To quantify deviations from this relation we define~\footnote{For model 10$(\mu)$ we use instead $\bigg\vert 1+\frac{A_{\text{CP}}(D^0  \to \pi^+ \pi^-)}{A_{\text{CP}}(D^0 \to K^+ K^-)} \bigg\vert$ to avoid  $\text{U}^{\text{tot}}_{\text{break}} >1$. It is tacitly understood that $K, Q_2,d_2$ and $\pi, Q_1,d_1$--indices in Eq.~\eqref{eq:UspinGen} and following need to be swapped in this case. \label{foot:swap}}
\begin{align}
\begin{split}\label{eq:UspinEstim}
    \text{U}^{\text{tot}}_{\text{break}} &= \bigg\vert 1+\frac{A_{\text{CP}}(D^0  \to K^+ K^-)}{A_{\text{CP}}(D^0 \to \pi^+ \pi^-)} \bigg\vert~.
\end{split}
\end{align}
In the U--spin limit $\text{U}^{\text{tot}}_{\text{break}}= 0$.

Using Eqs.~\eqref{eq:acp_zprime}, $\text{U}^{\text{tot}}_{\text{break}}$ can be written as
\begin{align}\label{eq:UspinGen}
\text{U}^{\text{tot}}_{\text{break}}&=\left\vert 1+ \frac{c_K\,F_{Q_2} + d_K\,F_{d_2}}{c_\pi\,F_{Q_1} + d_\pi\,F_{d_1}} \right\vert\,.
\end{align}

\begin{table}[ht!]
\renewcommand*{\arraystretch}{1.2}
 \centering
 \begin{tabular}{cccccccc}
  \hline
  \hline
 model & $\beta_9^{\mu\mu}$ & $\beta_{10}^{\mu\mu}$ & $\beta_9^{ee}$ & $\beta_{10}^{ee}$ & $\beta_{\pi^0}$ & $\beta_{\pi^\prime}$ & $\text{U}^{\text{tot}}_{\text{break}}\,$   \\
  \hline
$2$ & $\phantom{-}0.57$ & $-0.57$ & $-0.68$ & $\phantom{-}0.68$ & $-0.02$ & $-0.02$ & $0.42$\\
$4$ & $-1.04$ & $-0.35$ & $\phantom{-}1.04$ & $\phantom{-}0.35$ & $-0.03$ & $-0.03$ & $0.22$\\
$5$ & $-0.67$ & $\phantom{-}0$ &  $\phantom{-}0.67$ & $\phantom{-}0$ & $-0.10$ & $-0.09$ & $0.32$\\
$9$ & $-20.56$ & $-14.07$ & $\phantom{-}15.15$ & $-2.17$ & $-1.89$ & $-1.75$ & $0.22$\\
$10$ & $\phantom{-}37.25$ & $\phantom{-}3.39$ & $-32.73$ & $\phantom{-}1.13$ & $\phantom{-}1.31$ & $\phantom{-}1.22$ & $0.91$ \\
$10\mu$ & $-4.52$ & $-4.52$ & $-32.73$ & $\phantom{-}1.13$ & $\phantom{-}1.31$ & $\phantom{-}1.22$ & $0.91$ \\
\hline
\hline
  \end{tabular}
   \caption{Values of $\beta_{9/10}^{\ell\ell}$  in $({\rm TeV})^{-2}$ for $\ell=\mu,e$ and dimensionless $\beta_{\pi^{\prime,0}}$ as defined in Eq.~\eqref{eq:betapm} and Eq.~\eqref{eq:betapiprime}, respectively, as well as $\text{U}^{\text{tot}}_{\text{break}}$ in Eq.~\eqref{eq:UspinGen}, see footnote~\ref{foot:swap}, for $M_{Z^\prime}= 6\,\TeV$. }
   \label{tab:betavalues}
\end{table}

In TABLE~\ref{tab:betavalues} we give $\text{U}^{\text{tot}}_{\text{break}}\,$ for models $2$, $4$, $5$, $9$ and $10(\mu)$, for $M_{Z^\prime}=6\,\TeV$. 
The variation of $\text{U}^{\text{tot}}_{\text{break}}\,$ with $M_{Z^\prime}$ in the range shown is within a few percent.

Taking advantage of the smallness of the parameters $d_{K,\pi}$  relative to $c_{K,\pi}$, we  perform a Taylor expansion in Eq.~\eqref{eq:UspinGen} up to $\mathcal{O}(d_{K},d_{\pi})$  to qualitatively understand how U--spin breaking in our models emerges. This leads to
\begin{align}\label{eq:approxuspin}
    \text{U}^{\text{tot}}_{\text{break}}\approx\bigg\vert 1 + \frac{c_K}{c_\pi} - \frac{c_K d_\pi F_{d_1}}{c_\pi^2 F_{Q_1}} + \frac{d_K F_{d_2}}{c_\pi F_{Q_1}}\bigg\vert~,
\end{align}
for $F_{Q_1}=F_{Q_2}\neq 0$ (models 2, 4 and  5), while for $F_{Q_1}=F_{Q_2}= 0$ (models 9 and 10($\mu$)) Eq.~\eqref{eq:UspinGen} simply becomes
\begin{align}\label{eq:approxuspin0}
\text{U}^{\text{tot}}_{\text{break}}&=\left\vert 1+ \frac{d_K\,F_{d_2}}{d_\pi\,F_{d_1}} \right\vert\,.
\end{align}

\begin{table}
 \centering
 \begin{tabular}{lccc}
 \hline
 \hline
\quad$\times 10^{-4}$         & Data  & $\sigma_{\text{LHCb}}$  & $\sigma_{\text{Belle II}}$  \\
  \hline
$  \Delta A_{\text{CP}}$ &    $-15.4 \pm 2.9$~\cite{Aaij:2019kcg}  & $1.3\,(0.3)$  & -- \\
$  \Delta A^{\text{HFLAV}}_{\text{CP}}$ &    $-16.4 \pm 2.8$~\cite{Amhis:2019ckw}  & $1.3\,(0.3)$  & -- \\
$  A_{\text{CP}}(D^0 \to K^+ K^-)$ &  $-9  \pm 11$~\cite{Amhis:2019ckw}  & $3\,(0.7)$&  $3$ \\
$A_{\text{CP}}(D^0 \to \pi^+  \pi^-)$ & $-1 \pm 14$~\cite{Amhis:2019ckw} & $3\,(0.7)$& $5$ \\
$A_{\text{CP}}(D^0 \to \pi^0  \pi^0)$ & $-3 \pm 64$~\cite{Amhis:2019ckw} & -- & $9$ \\
$A_{\text{CP}}(D^+ \to \pi^+  \pi^0)$ &  $+290 \pm 290 \pm 30$~\cite{Mendez:2009aa}  & -- & $17$ \\
\hline
\hline
   \end{tabular}
   \caption{CP--asymmetries and future sensitivities  $\sigma$ in units of $10^{-4} $ at LHCb  Run 1-3  (Run 1-5)  \cite{Bediaga:2018lhg}  and Belle II with 50 $\text{ab}^{-1}$ \cite{Kou:2018nap}. }
   \label{tab:CP--data}
\end{table}

For models with $F_{Q_1}=F_{Q_2}\neq 0$ different sources of U--spin breaking exist. The second term in Eq.~\eqref{eq:approxuspin} accounts for effects originating from  interference between the SM--amplitude and the $F_{Q_{1,2}}$--charges. 
This contribution is responsible for  $22\,$\% U--spin breaking, which is of  the same order of magnitude as the expected U--spin breaking uncertainty of the SM. 
In contrast, the last two terms in Eq.~\eqref{eq:approxuspin} are pure NP U--spin breaking effects. Eq.~\eqref{eq:approxuspin} can further be simplified with $d_K\approx\frac{c_K}{c_\pi}\,d_\pi$ due to $\chi_\pi\approx\chi_K$, which holds numerically at the level of $\mathcal{O}(0.1-1)\,\%$. It follows that 
\begin{align}\label{eq:approxuspin1}
    \text{U}^{\text{tot}}_{\text{break}}\approx\bigg\vert 1 + \frac{c_K}{c_\pi} +\frac{d_K}{c_\pi}\bigg(\frac{F_{d_2}-F_{d_1}}{F_{Q_1}}\bigg)\bigg\vert   \, , 
\end{align}
highlighting that pure NP U--spin breaking effects are induced by  
\begin{align}\label{eq:approxuspin2}
    \text{U}^{\text{NP}}_{\text{break}}= \frac{d_K}{c_\pi} \bigg\vert \frac{F_{d_2}-F_{d_1}}{F_{Q_1}} \bigg\vert \approx 0.04\,\bigg\vert \frac{F_{d_2}-F_{d_1}}{F_{Q_1}}\bigg \vert~,
\end{align}
which indicates how both the pion chiral enhancement and $r_2$ suppress U--spin breaking in these models. Therefore, values of $F_{d_2}-F_{d_1}\sim\mathcal{O}(1)$ such as in model 5, induce  U--spin breaking within the range expected within the SM $\lesssim  30\,$\%.
In model 4, $F_{d_i}=0$ and 
$ \text{U}^{\text{NP}}_{\text{break}} = 0$, that is, U--spin breaking is SM-like. On the other hand,  for $F_{d_2}-F_{d_1}\sim\mathcal{O}(10)$ as in model 2, large U--spin breaking effects can arise and would be discernible with future sensitivities for $A_{\text{CP}}(K^+K^-)$ and  $A_{\text{CP}}(\pi^+\pi^-)$ shown in TABLE~\ref{tab:CP--data}.

For models with $F_{Q_1}=F_{Q_2}=0$ we obtain for the pure NP U--spin breaking from Eq.~\eqref{eq:approxuspin0}
\begin{align}\label{eq:approxuspinF0}
    \text{U}^{\text{NP}}_{\text{break}}(F_{Q_{1,2}}=0) \approx 0.78\,\bigg \vert  \frac{F_{d_2}-F_{d_1}}{F_{d_1}} \bigg  \vert \, , 
\end{align}
which, unlike in Eq.~\eqref{eq:approxuspin2}, is unsuppressed. Models with $F_{Q_1}=F_{Q_2}=0$ are therefore prime candidates for sizable NP U--spin breaking effects.
Models 9 and 10($\mu$) have been constructed for this purpose. However, in model 9
$F_{d_2}=F_{d_1}$ and U--spin breaking arises from $d_K \neq -d_\pi$ only, and is SM-like. 

Note, the strong phases associated with NP are assumed to be similar, $\sin \delta_\pi \simeq \sin \delta_K$, and order one; violation of Eq.~\eqref{eq:UB1} can be suppressed or even further enhanced by U--spin breaking in the strong phases. While this is an uncertainty on the NP interpretation, $Z^\prime$--signals could even be more striking.

In FIGs.~\ref{fig:uspin} and \ref{fig:uspinfuture} we show the contributions of models 2, 5, 9 and 10($\mu$) to the individual CP--asymmetries $A_{\text{CP}}(K^+K^-)$ and  $A_{\text{CP}}(\pi^+\pi^-)$ in blue, magenta, yellow and cyan, respectively. The U--spin limit is given by the red dashed line with $30\,\%$ U--spin breaking indicated by the red contour. Present experimental bounds from TABLE~\ref{tab:CP--data} are shown in FIG.~\ref{fig:uspin} as $1\sigma$ regions in gray for the individual asymmetries and in green for $\Delta A_{\text{CP}}$. The future sensitivities are indicated in light (dark) gray and green bands in  FIG.~\ref{fig:uspinfuture} for LHCb Run 1-3 (1-5). 
We use the following central values for the plot to the left (right)
\begin{align}
\label{eq:centralvals}
\begin{split}
  A_{\text{CP}}^{{\text{cen}}}(K^+ K^-) &= -0.6\cdot 10^{-3}\,\left(-1.45\cdot 10^{-3}\right)~,\\  
  A_{\text{CP}}^{{\text{cen}}}(\pi^+ \pi^-)&= \phantom{-}1.0\cdot 10^{-3}\,\left(\phantom{-}0.15\cdot 10^{-3}\right)~.
\end{split}
\end{align}
The orange error ellipses illustrate the NP sensitivity of the projected uncertainties of $A_{\text{CP}}(K^+K^-)$ and  $A_{\text{CP}}(\pi^+\pi^-)$  assuming no correlations. 
A future data-based analysis which takes into account correlations between the individual asymmetries and $\Delta A_{\text{CP}}$ can be expected to be more powerful.

U--spin symmetry within the SM is broken at the level of $30\,\%$. We find that flavorful $Z^\prime$--models can exceed this  by far (model 10($\mu$)), or moderately (model 2), which makes the measurements of  $A_{\text{CP}}(K^+K^-)$ and  $A_{\text{CP}}(\pi^+\pi^-)$ smoking guns for NP, within reach of  Belle II and LHCb with the projected sensitivities.

\subsection{Isospin breaking patterns in $\texorpdfstring{\boldsymbol{D^+ \to \pi^+ \pi^0}}{D+ to Pi+Pi0}$}\label{sec:I}

Isospin breaking arises in $Z^\prime$--models if $F_{u_1} \neq F_{d_1}$. In charm physics, the hadronic decay $D^+\to\pi^+\pi^0$ represents a formidable candidate to study these effects, because the CP--asymmetry $A_{\text{CP}} (\pi^+ \pi^0)$, defined by
\begin{align}
A_{\text{CP}}(\pi^+\pi^0)=\frac{\Gamma(D^+\to f^+)-\Gamma(D^-\to f^-)}{\Gamma(D^+\to f^+)+\Gamma(D^-\to f^-)}~,
\end{align} 
with $f^\pm=\pi^\pm \pi^0$ is a clean SM null test \cite{Grossman:2012eb}. 

Following the same procedure as in Section~\ref{sec:DACP} for $\Delta A_{\text{CP}}^{\text{NP}}$ we obtain, using $\theta_u \ll1$,
\begin{align}\label{eq:.acppmp0}
A_{\text{CP}}^{\text{NP}} (\pi^+ \pi^0)  \sim \frac{g_4^2}{M_{Z^\prime}^2} \,\theta_u\,  \Delta F_R\,  d_{\pi^\prime}\, (F_{d_1} - F_{u_1})~,
\end{align}
with 
\begin{align}
    d_{\pi^\prime}= -\frac{1}{a_{\pi^\prime}}\,r_2(m_c,M_{Z^\prime})~.
    \label{eq:dpiprimedef}
\end{align}
Here, $a_{\pi^\prime}$ denotes the tree-level contribution to $D^+\to\pi^+\pi^0$ whose modulus has been fixed experimentally, see Appendix~\ref{app:fit} for details. Numerical values of $d_{\pi^\prime}$ for different values of $M_{Z^\prime}$ are given in TABLE~\ref{tab:constants_acp}. 
Inserting Eq.~\eqref{eq:acp_zprime_formula} into Eq.~\eqref{eq:.acppmp0}, we obtain
\begin{align} \label{eq:+0}
  A_{\text{CP}}^{\text{NP}} (\pi^+ \pi^0)  \sim    \beta_{\pi^\prime}\cdot\Delta A_{\text{CP}}^{\text{NP}}  ~,
\end{align}
where
\begin{align}\label{eq:betapiprime}
    \beta_{\pi^\prime} =\frac{d_{\pi^\prime}\,(F_{d_1}-F_{u_1})}{c_K\,F_{Q_2}+d_K\,F_{d_2}-c_\pi\,F_{Q_1}-d_\pi\,F_{d_1}}~.
\end{align}
Values of $\beta_{\pi^\prime}$ for  $M_{Z^\prime}= 6\,\TeV$ and different $Z^\prime$--models can be seen in TABLE~\ref{tab:betavalues}.
Since we have lost information about the signs of the leading SM decay amplitudes
with which NP is interfering, we
cannot predict the relative sign between the CP--asymmetries in Eq.~\eqref{eq:+0} without relying on assumptions on the strong interaction.
Note, unlike for $A_{\text{CP}}(K^+K^-)$ and  $A_{\text{CP}}(\pi^+\pi^-)$, there is no SM flavor symmetry here at work.

We find that model 9 and 10$(\mu)$ induce values near
\begin{align}  A_{\text{CP}}^{{\text{NP}}} (\pi^+ \pi^0)  \sim (1 -2)\cdot  \Delta A_{\text{CP}}^{{\text{NP}}} \, ,
\end{align} which  for $\Delta A_{\text{CP}}^{{\text{NP}}} \sim 10^{-3}$ is within  the projected sensitivity of  Belle II  with 
50 ${\rm ab}^{-1}$~\cite{Kou:2018nap}, see TABLE~\ref{tab:CP--data}.
Model 2, 4 and 5 induce $ A_{\text{CP}}^{\text{NP}} (\pi^+ \pi^0)  \lesssim 0.1 \cdot \Delta A_{\text{CP}}^{\text{NP}} \sim 10^{-4}$, beyond the reach of current facilities.

This behavior can be understood by expanding Eq.~\eqref{eq:betapiprime} in the $d_i$ up to $\mathcal{O}(d_i)$. For $F_{Q_1}=F_{Q_2}=0$ (model 9 and 10{($\mu$)}), we find that $\beta_{\pi^\prime}$ scales with $d_{\pi^\prime}/d_K\approx {-1.6}$ times a combination of charges $(F_{d_1}-F_{u_1})/F_{d_2} (1+...)\sim \mathcal{O}(1)$ resulting in  $\mathcal{O}(1)$ isospin breaking effects.
For models with $F_{Q_1}=F_{Q_2}\neq 0$ instead a suppression factor $d_{\pi^\prime}/(c_K-c_\pi)\approx 0.03$ exists from the chiral enhancement of the $(V-A)\times(V+A)$ operators, leading to $\beta_{\pi^\prime}$  of $\mathcal{O}(10^{-2}-10^{-1})$. 

\subsection{\texorpdfstring{$\boldsymbol{D^0 \to \pi^0 \pi^0}$}{D0 to pi0pi0} \label{sec:pi0}}

We work out the CP--asymmetry for $D^0\to\pi^0\pi^0$ decays because of its potential to diagnose patterns of NP~\cite{Hiller:2012xm}.
In addition, the experimental prospect at Belle II for $A_{\text{CP}}(D^0\to\pi^0\pi^0)$ is about a factor of two better than for $A_{\text{CP}}(D^+\to\pi^+\pi^0)$, see TABLE~\ref{tab:CP--data}.
In the $Z^\prime$--models, 
$A_{\text{CP}}(D^0\to\pi^0\pi^0)$ is obtained  from Eqs.~\eqref{eq:.acppmp0} and~\eqref{eq:dpiprimedef} after replacing  subscripts ${\pi^\prime}$ by ${\pi^0}$
with otherwise identical expressions.
Therefore, with $\beta_{\pi^0}$ given in TABLE~\ref{tab:betavalues},  
\begin{align} \label{eq:00}
A_{\text{CP}}^{\text{NP}} (\pi^0 \pi^0)   \sim  \beta_{\pi^0}{\cdot} \Delta A_{\text{CP}}^{\text{NP}}  ~,
\end{align}
hence 
\begin{align} A_{\text{CP}}^{{\text{NP}}} (\pi^0 \pi^0) \lesssim 2\cdot \Delta A_{\text{CP}}^{{\text{NP}}} \, ,
\end{align} 
with the limit saturated by model 9, and which is within the sensitivity of Belle II with 
50 ${\rm ab}^{-1}$~\cite{Kou:2018nap}, see TABLE~\ref{tab:CP--data}.
Furthermore,
\begin{align} \label{eq:pions}
\frac{  A_{\text{CP}}^{\text{NP}} (\pi^0 \pi^0)  }{  A_{\text{CP}}^{\text{NP}} (\pi^+ \pi^0)} \sim \frac{  \beta_{\pi^0}}{ \beta_{\pi^\prime}} = \frac{  a_{\pi^\prime}}{ a_{\pi^0}} = 1.08 \pm 0.10\, ,
\end{align}
holds universally for all $Z^\prime$--models with $F_{u_1} \neq F_{d_1}$.
Experimental tests of Eq.~\eqref{eq:pions} can support a  $Z^\prime$--interpretation, however, additional uncertainties from large, unknown strong phases exist, which can modify the relation.
As discussed after Eq.~\eqref{eq:+0}, we cannot predict the relative sign between the CP--asymmetries \eqref{eq:00}, \eqref{eq:pions} without relying on input on the strong interaction.

\section{Semileptonic decays vs. \texorpdfstring{$\boldsymbol{\Delta A_{\text{CP}}}$}{Delta ACP}   \label{sec:slCP}}

The dominant Wilson coefficients in $c\to u \,\ell^+\ell^-$ transitions are $C_{9/10}^{\ell\ell\,(\prime)}$, defined in Eq.~\eqref{eq:Heffsl}. In flavorful $Z^\prime$--models~\cite{Bause:2019vpr}
\begin{align}
  C_{9/10}^{\ell\ell\,}\left(M_{Z^\prime}\right)\,& =\,-\,\frac{\pi}{\sqrt{2}\,G_F\,\alpha_e}\,\frac{g_L^{uc}}{M_{Z^\prime}^2}\,\left(g_R^{\ell\ell}\pm g_L^{\ell\ell}\right)~, \\
  C_{9/10}^{\ell\ell\,\prime}\left(M_{Z^\prime}\right)\,&=\,-\,\frac{\pi}{\sqrt{2}\,G_F\,\alpha_e}\,\frac{g_R^{uc}}{M_{Z^\prime}^2}\,\left(g_R^{\ell\ell}\pm g_L^{\ell\ell}\right)~,\label{eq:WCsemileptonicsZp}
\end{align}
where $g_{R}^{\ell\ell}\,=\,g_4\,F_{e_i}$ and $g_{L}^{\ell\ell}\,=\,g_4\,F_{L_i}$ with in general different couplings for muons and electrons.
As explained in Section~\ref{sec:DDbar}, we analyze in this work $Z^\prime$--models with $g^{uc}_L=0$ and  Im$(g^{uc}_R)$ large.

CP--asymmetries in the branching ratios are induced by interference of NP, here through $g^{uc}_R$, with $C_9^{\rm eff}$, the effective coefficient of $O_9$ present in the SM, which is lepton-universal, depends on the dilepton invariant mass and has sizable hadronic contributions and provides sizable strong phases. 
This interference term is sensitive to $C_{9}^{\ell\ell\,\prime}$ only.
Angular analysis offers further opportunities. 
An interesting recent example for the latter is $D^0 \to \pi^+ \pi^- \mu^+ \mu^-$ decays ~\cite{Aaij:2018fpa,deBoer:2018buv,Cappiello:2012vg}.
Notably, the angular observables $I_{5,6,7}$ are GIM--protected in the SM and clean null tests ~\cite{deBoer:2018buv}.
In the $Z^\prime$--models under consideration, $I_{5,6}$ are induced by $\text{Re}(C_9^{\ell\ell\,\prime} \cdot C_{10}^{\ell\ell\,\prime *})$ and $\text{Im}(C_{10}^{\ell\ell\,\prime} \cdot C_{9}^{{\rm eff} *})$, whereas $I_7$ is induced by $\text{Re} \left[ (C_{9}^{\text{eff} *}-C_9^{\ell\ell\,\prime}) \cdot C_{10}^{\ell\ell\,\prime *}) \right]$.
CP--asymmetries in angular asymmetries, on the other hand, can stem from naïve $T$--odd observables and do not rely on strong phases ($I_{7,8,9}$).
CP--odd ones ($I_{5,6,8,9}$) provide CP--asymmetries that can be measured without tagging, see Ref.~\cite{deBoer:2018buv} for details.
A complete and detailed  analysis of angular asymmetries in $Z^\prime$--models is beyond the scope of this work.
What we do want to point out here is that a global analysis of angular and CP--asymmetries can probe both $C_{9}^{\ell\ell\,\prime}$ and  $C_{10}^{\ell\ell\,\prime}$ for 
electrons, $\ell=e$ and muons, $\ell=\mu$ separately, and therefore
can distinguish different $U(1)^\prime$--charge assignments.

\begin{figure*}[!t]
    \centering
\includegraphics[width=0.45\textwidth]{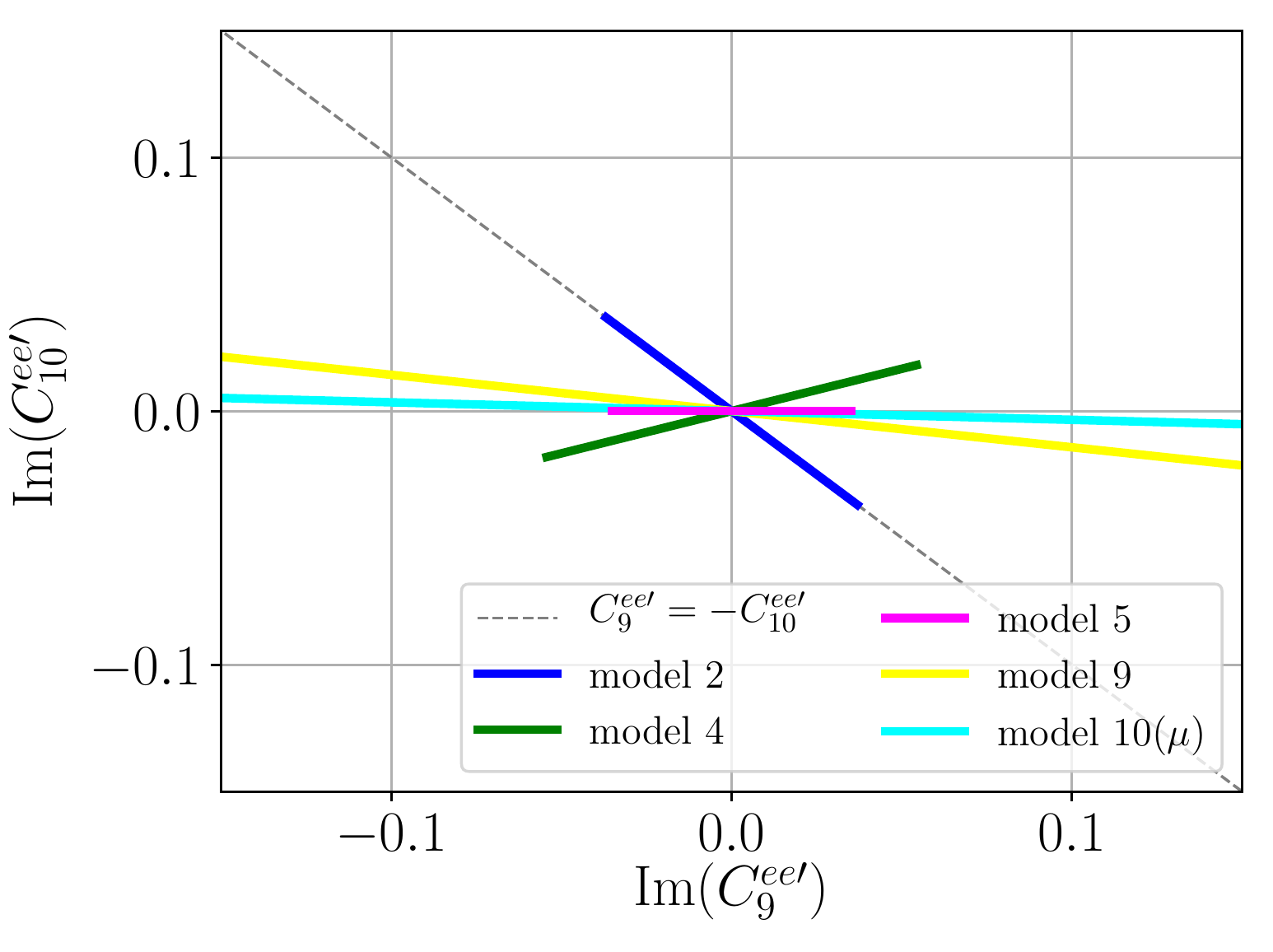}
\includegraphics[width=0.45\textwidth]{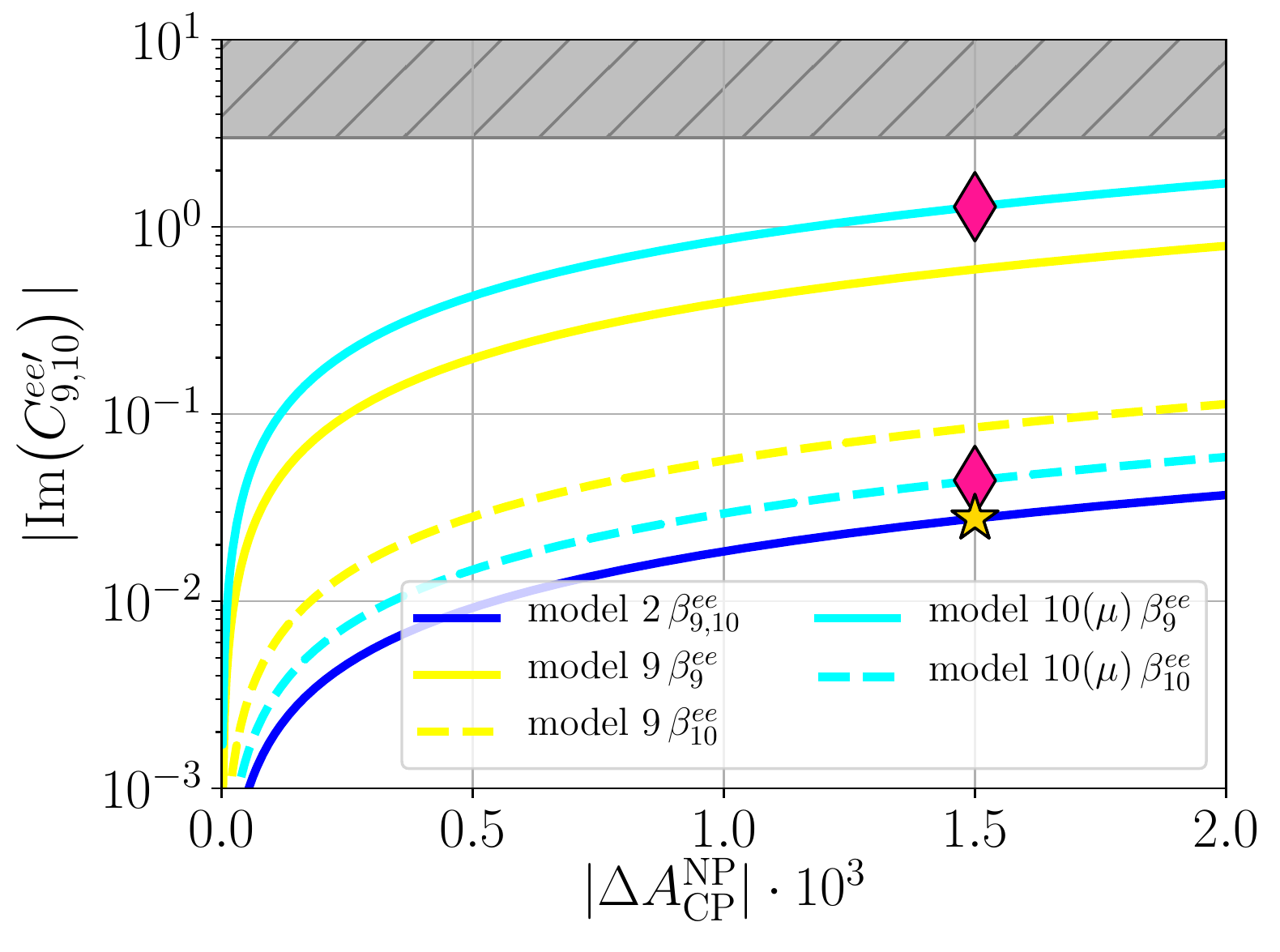}
\includegraphics[width=0.45\textwidth]{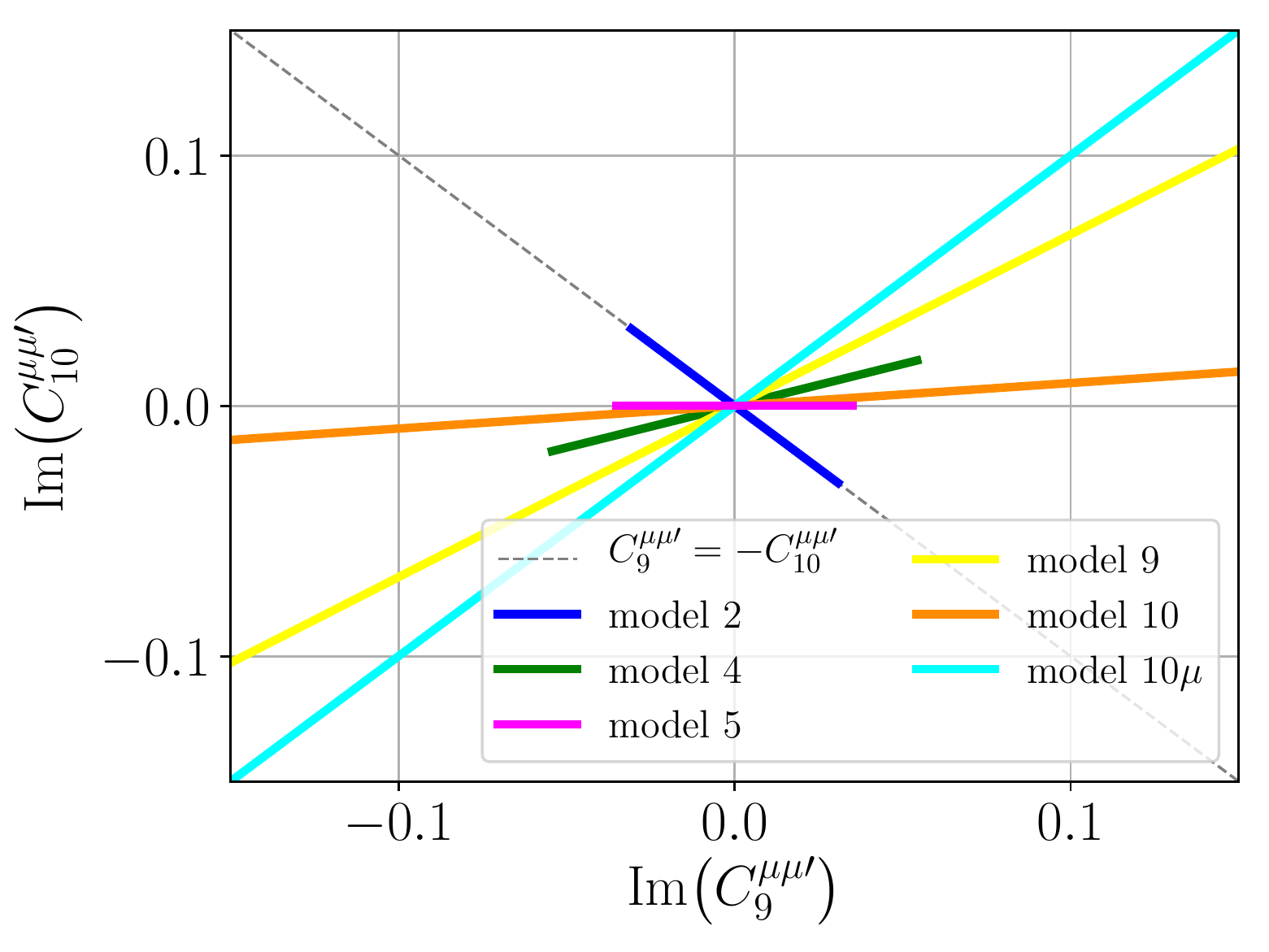}
\includegraphics[width=0.45\textwidth]{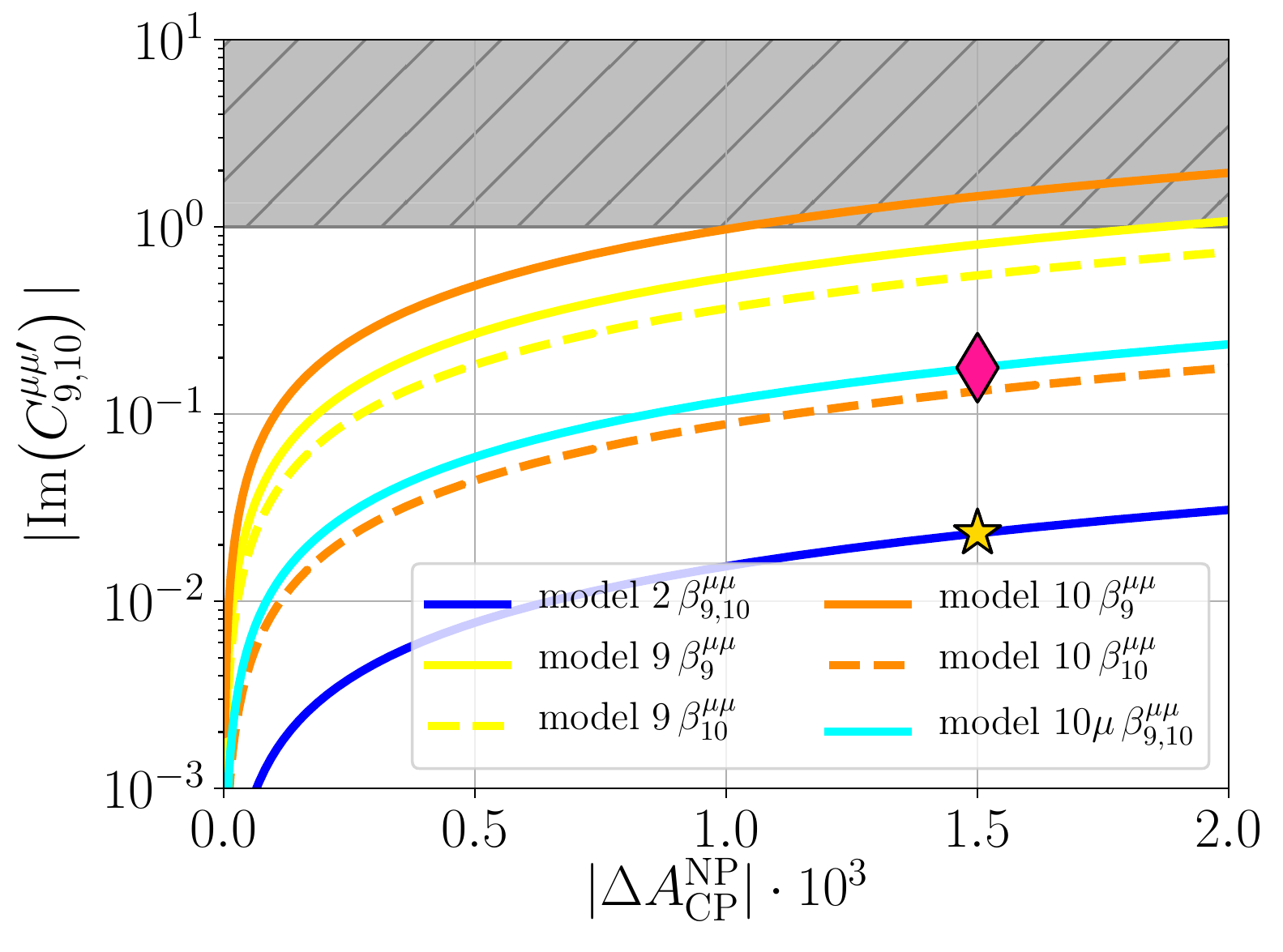}
\caption{The interplay between semi--electronic (upper plots) and semi--muonic (lower plots) charm FCNCs (\ref{eq:imc910})  and  $\Delta A_{\text{CP}}$. 
The lines for  model 2, 4 and 5 end when the corresponding $\Delta A_{\text{CP}}^{\text{NP}}$ exceeds $2 \cdot 10^{-3}$ (plots to the left).
In the plots to the right the correlation (\ref{eq:imc910}) between $  \text{Im}(C_{9}^{\ell\ell\,\prime})$ (solid) and  $  \text{Im}(C_{10}^{\ell\ell\,\prime})$ (dashed) 
and $\vert \Delta A_{\text{CP}}\vert$  in the $Z^\prime$--models 2, 9, 10 and 10$\mu$ is made explicit.
The golden star and pink diamond are benchmark points \eqref{eq:bench} (model 2) and \eqref{eq:bench10} (model 10$\mu$), respectively.
The shaded areas correspond to the upper limits~\eqref{eq:limits}. See text for details.}
\label{fig:sl}
\end{figure*}

Taking the imaginary part of Eq.~\eqref{eq:WCsemileptonicsZp} and employing Eq.~\eqref{eq:acp_zprime_formula}, we obtain
\begin{align} \label{eq:imc910}
    \text{Im}(C_{9/10}^{\ell\ell\,\prime}) \, \sim \,\frac{\pi}{\sqrt{2}\,G_F\,\alpha_e}\,  \beta_{9/10}^{\ell \ell} \cdot \Delta A_{\text{CP}}^{\text{NP}}~,
\end{align}
where 
\begin{align}
\label{eq:betapm}
    \beta_{9/10}^{\ell \ell}= \frac{F_{e_i}\pm F_{L_i}}{c_K\,F_{Q_2} + d_K\,F_{d_2}-c_\pi\,F_{Q_1} - d_\pi\,F_{d_1}} ~.
\end{align}
Values of $\beta_{9/10}^{\ell \ell}$ for $\ell=\mu,e$ in $({\rm TeV})^{-2}$  are given in TABLE~\ref{tab:betavalues}. For $\Delta A_{\text{CP}}^{\text{NP}}\sim 10^{-3}$ we find
\begin{equation}\label{eq:estimate_c910_acp}
    \text{Im}(C_{9/10}^{\ell\ell\,\prime})\,\sim\,0.03  \, ({\rm TeV})^{2} \cdot  \beta_{9/10}^{\ell \ell}  \,,
\end{equation}
consistent with $C_{9/10}^{\ell\ell\,\prime}=\mathcal{O}(10^{-2})$ for $g_L^{uc}=0, g_R^{uc} \neq 0$~\cite{Bause:2019vpr} and for 
$\beta_{9/10}^{\ell \ell}=\mathcal{O}(1/{\rm TeV}^{2})$ (models 2, 4  and 5).
Models 9 and 10($\mu$)  have sizable couplings to leptons, and in addition $F_{Q_{1,2}}=0$, which bring a factor of $c_{\pi,K}/d_{\pi,K}$, see Eq.~\eqref{eq:betapm},
score $\beta_{9/10}^{\ell \ell}=\mathcal{O}(10/{\rm TeV}^{2})$ and sizable $C_{9/10}^{\ell\ell\,\prime}=\mathcal{O}(10^{-1})$.
As values of $ \text{Im}(C_{9/10}^{\ell\ell\,\prime}) \gtrsim \mathcal{O}(10^{-2}-10^{-1})$ suffice  to  induce CP--asymmetries beyond the SM in semileptonic $D$--decays at the few percent level and above~\cite{Fajfer:2012nr,deBoer:2015boa,deBoer:2018buv,Bause:2019vpr}, all models can simultaneously lead to  $\Delta A_{\text{CP}}^{\text{NP}}\sim 10^{-3}$ with NP patterns in $c \to u\, \ell^+ \ell^-$ decays.

In FIG.~\ref{fig:sl} we show the imaginary part of Wilson coefficients with di-electrons (upper plots) and di-muons (lower plots)
for different models as in Eq.~(\ref{eq:imc910}). 
Plots to the left show lepton vector couplings versus lepton axial vector couplings, $  \text{Im}(C_{9}^{\ell\ell\,\prime})$ vs.~$  \text{Im}(C_{10}^{\ell\ell\,\prime})$, respectively.
Also given is  $\text{Im}(C_{10}^{\ell\ell\,\prime})=-\text{Im}(C_{9}^{\ell\ell\,\prime})$ (thin gray line).
The lines corresponding to model 2, 4, and 5 end when the corresponding $\Delta A_{\text{CP}}^{\rm NP}$ exceeds $2 \cdot 10^{-3}$.
Results are lepton non-universal as anticipated and
sensitive to the lepton doublet and singlet charges.
In the plots to the right the correlation~\eqref{eq:imc910} between $  \text{Im}(C_{9}^{\ell\ell\,\prime})$ (solid) and  $  \text{Im}(C_{10}^{\ell\ell\,\prime})$ (dashed) 
and $\vert \Delta A_{\text{CP}}^{\text{NP}}\vert$  in the $Z^\prime$--models 2, 9, 10 and 10$\mu$ is made explicit. Curves for models 4 and 5 are only in mild excess of those for model 2, or smaller, see TABLE~\ref{tab:betavalues}, and  are not shown to avoid clutter.

As couplings to electrons and muons differ, lepton non-universality in charm~\cite{Fajfer:2015mia, deBoer:2018buv,Bause:2019vpr} is induced, for example 
in the ratio of branching ratios of $D \to \pi\, \mu^+ \mu^-$ and
$D \to \pi\, e^+ e^-$ using identical kinematic cuts, $R_\pi^D$. To better control SM backgrounds from intermediate resonances $R=\phi, \eta^{(\prime)}, \rho,\ldots$, via $D \to \pi R (\to  \ell^+ \ell^-)$, interesting regions are for low (high) dilepton mass, below the $\eta$--mass (above the $\phi$--mass), see Ref.~\cite{Bause:2019vpr} for details.
We focus on the high mass region as it has fewer sensitivity to unknown strong phases from the resonances.

Using $\beta_{9/10}^{\ell\ell}$ from TABLE ~\ref{tab:betavalues} and Eq.~\eqref{eq:estimate_c910_acp} we find that all models yield order one deviations from the universality limit $R_\pi^D=1$.
Except for model 10$\mu$, which has smaller couplings to muons by construction, all models can induce
significant enhancements or suppressions from the SM.
In particular, in the high mass region, for $\phi_R=\pi/2$ and varying strong resonance phases, see Ref.~\cite{Bause:2019vpr} for details,
\begin{align}
\begin{split}
R_\pi^D &  \sim  \lbrack 0.6 \ldots 1.5 \rbrack \quad (\text{model}\,2,4,5)  ~,\\
R_\pi^D & \sim  \lbrack 0.2 \ldots 70 \rbrack \quad   (\text{model}\,9)  ~,\\
R_\pi^D & \sim  \lbrack 0.2 \ldots 11 \rbrack \quad  (\text{model}\,10) ~, \\
R_\pi^D & \sim  \lbrack 0.03 \ldots 0.8 \rbrack \quad  (\text{model}\,10\mu)  ~,
\end{split}
\end{align}
allowing to signal NP.

\section{Conclusions}\label{sec:con}

Patterns of observables are indispensable for pinning down an underlying NP--dynamics. We looked globally  into
hadronic and semileptonic charm decays and their respective CP--asymmetries.
We find that there is strong benefit in doing so.

Most important, all flavorful, anomaly-free $Z^\prime$--models in TABLE~\ref{tab:ZprimeModels} can simultaneously accommodate $\Delta A_{\text{CP}}^{\text{NP}} \sim 10^{-3}$
and induce measurable CP--asymmetries in the semileptonic  $c \to u \, \ell^+ \ell^-$ modes  for $\ell=e$ or $\ell=\mu$ above the SM.
An observation of CP--violation in, for instance,  $D \to \pi \, \ell^+ \ell^-$ or  $D \to \pi \pi\, \ell^+ \ell^-$ decays supports a NP--interpretation of $\Delta A_{\text{CP}}$, Eqs.~\eqref{eq:deltaACP} and \eqref{eq:worldaverage}, see FIG.~\ref{fig:sl}.

Additional cross checks are provided  by CP--asymmetries in $D^0 \to \pi^+ \pi^-$,  $D^0 \to K^+ K^-$, which probe for  U--spin breaking NP, see FIGs.~\ref{fig:uspin} and \ref{fig:uspinfuture} for present data and future sensitivities, respectively. In addition, isospin violating NP can be observed with projected sensitivities at Belle II
in $D^0 \to \pi^0 \pi^0$, $D^+ \to \pi^+ \pi^0$ decays, whose CP--asymmetries can exceed $\Delta A_{\text{CP}}$, Eqs.~\eqref{eq:+0} and \eqref{eq:00}.

In the $Z^\prime$--models lepton non--universality is generic, and observable in the ratio of  branching fractions of $D \to \pi\, \mu^+ \mu^-$ and
$D \to \pi\, e^+ e^-$ decays, as briefly discussed in Section~\ref{sec:slCP}. 
The $Z^\prime$--model 9 with order one enhancements over the universality limit, $R_\pi^D \gg 1$, also
induces $A_{\text{CP}}^{{\text{NP}}} (\pi^+ \pi^0) \sim  A_{\text{CP}}^{{\text{NP}}} (\pi^0 \pi^0) \lesssim 2\cdot \Delta A_{\text{CP}}^{\text{NP}}$.
 $Z^\prime$--model 10$\mu$ with order one suppression of the universality limit, $R_\pi^D<1$ exhibits sizable NP U--spin breaking
$A_{\text{CP}}^{{\text{NP}}} (\pi^+ \pi^-)  \ll  A_{\text{CP}}^{{\text{NP}}} (K^+ K^-) \sim \Delta A_{\text{CP}}$.

Checking correlations pins down models.
Improved data and sensitivities from LHCb and Belle II are important in this program.
We encourage and look forward to further CP--studies of rare semileptonic and hadronic charm decays.

\acknowledgments
This work is supported by the \textit{ Studienstiftung des Deutschen Volkes} (MG) and the \textit{ Bundesministerium f\"ur Bildung und Forschung} -- BMBF (HG).

\appendix 

\section{Experimental input}\label{app:fit}
 
We extract the modulus of the  dominant, SM decay amplitudes from data on branching ratios \cite{Tanabashi:2018oca}
given in TABLE~\ref{tab:a_factor_list}.
We use 
  \begin{align}
    \text{BR}(D\to P_1 P_2)=\frac{\vert\mathcal{A}_P\vert^2}{16\,\pi\, m_{D}}\,\sqrt{1-\frac{4\,m_P^2}{m_{D}^2}}\,\tau_{D}~,
\end{align}
where~\cite{Grossman:2006jg}
\begin{align}\label{eq:AmpDP}
    \mathcal{A}_P=\eta_P\,\lambda_P\,a_P\,\frac{G_F}{\sqrt{2}}\,\left(m_D^2-m_P^2\right)\,f_0^{D\to P}(m_P^2)\,f_P~,
\end{align}
$P=\pi, \pi^0, \pi^\prime, \,K$, $\lambda_{\pi}=\lambda_d$ and $\lambda_K=\lambda_s$ and
\begin{align}
    \eta_P=\begin{cases}
    1\quad &P=\pi,\,\pi^0,\,K \\
    \frac{1}{\sqrt{2}}\quad &P=\pi^\prime\\
    \end{cases}\,.
\end{align}
The subscript $\pi^\prime$ corresponds to the $D^+\to\pi^+\pi^0$ channel, and $\pi^0$ to $D^0\to\pi^0\pi^0$.
Relevant form factors $f_0^{D\to P}$ and decay constants $f_P$ are taken from Ref.~\cite{Lubicz:2017syv} and \cite{Tanabashi:2018oca}, respectively.
Resulting values of $a_P >0$ are given in TABLE~\ref{tab:a_factor_list}. 
\begin{table}[ht!]
\renewcommand*{\arraystretch}{1.5}
 \centering
 \begin{tabular}{lcc}
  \hline
  \hline
    mode & $\text{BR}\left(\text{mode}\right)$ & $a_P$ \\
  \hline
    $D^0\to K^+K^-$ & $(4.08\pm0.06)\cdot 10^{-3}$ & $1.19\pm0.04$ \\
    $D^0\to \pi^+\pi^-$ & $(1.455\pm0.024)\cdot 10^{-3}$ & $0.94\pm0.07$ \\
    $D^0\to \pi^0\pi^0$ & $(8.26\pm0.25)\cdot 10^{-4}$ & $0.71\pm0.05$ \\
    $D^+\to \pi^0\pi^+$ & $(1.247\pm0.033)\cdot 10^{-3}$ & $0.77\pm0.05$ \\
  \hline
  \hline
  \end{tabular}
   \caption{Measured branching ratios~\cite{Tanabashi:2018oca} and  $a_P$--parameters from Eq.~\eqref{eq:AmpDP} for different decay modes.}
   \label{tab:a_factor_list}
\end{table}

\section{Evolution of Wilson coefficients}\label{app:A}
 
The Wilson coefficients $\widetilde{C}_{7,8,9,10}^{(\prime)}$ at the $Z^\prime$ mass scale \eqref{eq:WCZp} are evolved to the charm mass scale at LO in $\alpha_s$.
The requisite anomalous dimension matrix for the operators ${\widetilde{Q}}_{7,8,9,10}$ can be inferred from Ref.~\cite{Buchalla:1995vs}.
We obtain
\begin{align}\label{eq:ADM}
\gamma^{0}_{F} = \begin{pmatrix}  
\frac{6}{N_C} & -6 & 0 & 0 \\
 0 & \frac{6(1-N_C^2)}{N_C} & 0 & 0 \\
 0 & 0 & \frac{-6}{N_C} & 6 \\
 0 & 0 & 6 & \frac{-6}{N_C}
\end{pmatrix},
\end{align}
where $N_C=3$ is the number of colors.
Since QCD conserves parity, $\gamma^0_F$ is identical for $\widetilde{Q}_i$  and  $\widetilde{Q}^\prime_i$.
Using Eq.~\eqref{eq:ADM}, the Wilson coefficients are evolved to the charm scale, integrating out  degrees of freedom at the ${(Z^\prime, t, b)}$--scales,
\begin{align*}
    \vec{C}(\mu)\,=\,U_4(\mu,m_b)\,\widehat{U}_5(m_b,m_t)\,\widehat{U}_6(m_t,M_{Z^\prime})\,\vec{C}(M_{Z^\prime})~,
\end{align*}
where $\widehat{U}_f (m_1,m_2) \equiv M_{f}(m_1)\,U_f(m_1,m_2)$ and  $U_f(m_1,m_2)$ is the evolution matrix from scale $m_2$ to scale $m_1$ in an effective field theory with $f$ active flavors; $M_f$ is  the threshold matrix that matches the two effective theories with $f-1$ and $f$ active flavors. At LO in $\alpha_s$, the $M_{f}$ matrices are equal to the identity matrix. For $\mu=m_c$ and  $M_{Z^\prime}=6\,$TeV, one finds
\begin{align}
\begin{split}
    \widetilde{C}^{(\prime)}_7(m_c)&= \phantom{-}0.829\,\widetilde{C}^{(\prime)}_7(M_{Z^\prime})~,\\
    \widetilde{C}^{(\prime)}_8(m_c)&= \phantom{-}1.224\,\widetilde{C}^{(\prime)}_7(M_{Z^\prime})+4.502\,\widetilde{C}^{(\prime)}_8(M_{Z^\prime})~,\\
    \widetilde{C}^{(\prime)}_9(m_c)&=\phantom{-}1.404\,\widetilde{C}^{(\prime)}_9(M_{Z^\prime})-0.718\,\widetilde{C}^{(\prime)}_{10}(M_{Z^\prime})~,\\
    \widetilde{C}^{(\prime)}_{10}(m_c)&=-0.718\,\widetilde{C}^{(\prime)}_9(M_{Z^\prime})+1.404\,\widetilde{C}^{(\prime)}_{10}(M_{Z^\prime})~.
    \end{split}
\end{align}
We use $m_c = (1.280\pm0.013)\,\GeV$, $m_b = (4.198\pm0.012)\,\GeV$~\cite{Aoki:2019cca} and $m_t=(165.9\pm2.1)\,\GeV$ ~\cite{Fuster:2017rev,Aad:2015waa} and central values for the thresholds.

\section{Hadronic matrix elements}\label{app:HME}

In order to estimate the NP decay amplitudes, we need to determine the hadronic matrix elements for each operator given by Eqs.~\eqref{eq:OpZ7}--\eqref{eq:OpZ10}. For that purpose, we employ factorization of currents, $P=\pi,K$,
\begin{align}
&\bra{P^+\,P^-} Q_i\ket{D^0}\\
&= \bra{P^+ }(\bar{q}_1\,\Gamma_1\, q_2)\ket{0}\,\bra{P^-}(\bar{q}_3\, \Gamma_2\, q_4)\ket{D^0}\,B^{P^+ P^-}_i~, \nonumber
\end{align}
where $Q_i=(\bar{q}_1\,\Gamma_1\,q_2)\,(\bar{q}_3\, \Gamma_2\, q_4)$ is a $4$--quark operator and $\Gamma_{1,2}$ represent possible Dirac and color structures while $q_j$ denote quarks. The factor $B^{P^+ P^-}_i$ parametrizes the deviation of the true hadronic matrix element from its naïve approximation, 
$B^{P^+ P^-}_i \vert_{\text{naïve}}=1$. For the NP contributions we work in this approximation.  
After employing Fierz identities in the flavor and color space, we find for $D^0\to K^+ K^-$ and $\pi^+\pi^-$ decays
\begin{align}
    \langle {\widetilde{Q}}_7 \rangle_{K,\pi}\,&=\,\,\frac{1}{N_C}\,\langle \widetilde{Q}_8 \rangle_{K,\pi}\,~,\label{eq:HME}\\
        \langle {\widetilde{Q}}_8 \rangle_{K,\pi}&= F_{d_2,\,d_1}\, \chi_{K,\pi}(\mu)\,\langle Q_1^{s,\,d} \rangle_{K,\pi}~,\\
        \langle {\widetilde{Q}}_9 \rangle_{K,\pi}\,&=\,\,\frac{1}{N_C}\,\langle {\widetilde{Q}}_{10} \rangle_{K,\pi}~,\\
        \langle {\widetilde{Q}}_{10} \rangle_{K,\pi}&=F_{Q_2,\,Q_1}\,\langle Q_1^{s,\,d} \rangle_{K,\pi}~,
\end{align}
where $\langle ...\rangle_{P}=\langle P^+ P^-|...|D^0\rangle$,  $Q_1^{p} = (\bar u p)_{V-A} (\bar p c)_{V-A}$ and $\chi_{K,\pi}(\mu)$ are the usual chiral enhancements generated by $(V-A)\times(V+A)$ operators,
\begin{align}
\begin{split}
    \chi_{K}(\mu) &= \frac{2\,M_K^2}{m_c(\mu)\,m_s(\mu)}~, \\ \chi_{\pi}(\mu) &= \frac{2\,M_\pi^2}{m_c(\mu)\,(m_d+m_u)(\mu)}~, 
\end{split}
\end{align}
with values $\chi_{K}(m_c) \approx 3.626$ and $\chi_{\pi}(m_c) \approx 3.655$ at the charm mass scale. For the $\widetilde{Q}_i^\prime$ operators the same relations hold but with the proper exchange of charges $F_{Q_i}\leftrightarrow F_{d_i}$.

For $D^+\to \pi^0\pi^+$ decays we find
\begin{align}
    \langle {\widetilde{Q}}_7 \rangle_{\pi^\prime}\,&=\,\,\frac{1}{N_C}\,\langle \widetilde{Q}_8 \rangle_{\pi^\prime}\,~,\\
        \langle {\widetilde{Q}}_8 \rangle_{\pi^\prime}&= \frac{\chi_{\pi}(\mu)}{\sqrt{2}}\,(F_{u_1}\,-\,F_{d_1})\,\langle Q_1^{u} \rangle_u~,\\
        \langle {\widetilde{Q}}_9 \rangle_{\pi^\prime}\,&=\,\,\frac{1}{N_C}\,\langle {\widetilde{Q}}_{10} \rangle_{\pi^\prime}=0~,\label{eq:HMEIB}
\end{align}
and for the corresponding $\widetilde{Q}_i^\prime$ operators 
\begin{align}
    \langle {\widetilde{Q}^\prime}_7 \rangle_{\pi^\prime}\,&=\,\,\frac{1}{N_C}\,\langle \widetilde{Q}^\prime_8 \rangle_{\pi^\prime}\,=\,0~,\\
        \langle {\widetilde{Q}^\prime}_9 \rangle_{\pi^\prime}\,&=\,\,\frac{1}{N_C}\,\langle {\widetilde{Q}^\prime}_{10} \rangle_{\pi^\prime}~,\\
        \langle {\widetilde{Q}^\prime}_{10} \rangle_{\pi^\prime}&= \frac{1}{\sqrt{2}}\,(F_{u_1}\,-\,F_{d_1})\,\langle {Q}_1^{u} \rangle_u~.\label{eq:HMEIBprime}
\end{align}
For $D^0\to \pi^0\pi^0$ decays we obtain
\begin{align}
    \langle {\widetilde{Q}}_7 \rangle_{\pi^0}\,&=\,\,\frac{1}{N_C}\,\langle \widetilde{Q}_8 \rangle_{\pi^0}\,~,\\
        \langle {\widetilde{Q}}_8 \rangle_{\pi^0}&= \frac{\chi_{\pi}(\mu)}{2}\,(F_{u_1}\,-\,F_{d_1})\,\langle Q_1^{u} \rangle_u~,\\
        \langle {\widetilde{Q}}_9 \rangle_{\pi^0}\,&=\,\,\frac{1}{N_C}\,\langle {\widetilde{Q}}_{10} \rangle_{\pi^0}=0~,\label{eq:HMEIBpipi}
\end{align}
and for the corresponding $\widetilde{Q}_i^\prime$ operators 
\begin{align}
    \langle {\widetilde{Q}^\prime}_7 \rangle_{\pi^0}\,&=\,\,\frac{1}{N_C}\,\langle \widetilde{Q}^\prime_8 \rangle_{\pi^0}\,=\,0~,\\
        \langle {\widetilde{Q}^\prime}_9 \rangle_{\pi^0}\,&=\,\,\frac{1}{N_C}\,\langle {\widetilde{Q}^\prime}_{10} \rangle_{\pi^0}~,\\
        \langle {\widetilde{Q}^\prime}_{10} \rangle_{\pi^0}&= \frac{1}{2}\,(F_{u_1}\,-\,F_{d_1})\,\langle {Q}_1^{u} \rangle_u~,\label{eq:HMEIBprimepipi}
\end{align}
where $\langle ...\rangle_{\pi^\prime}=\langle \pi^+ \pi^0|...|D^+\rangle$, $\langle ...\rangle_{\pi^0}=\langle \pi^0 \pi^0|...|D^0\rangle$ and $\langle ...\rangle_q=\langle \bar{q} q|...|D^+\rangle$. Eqs.~\eqref{eq:HME}--\eqref{eq:HMEIBprimepipi} are obtained in the isospin limit, $m_u=m_d$ and $e=0$, since these isospin breaking corrections from within the SM are negligible with respect to the NP ones, $F_{u_i,d_i,Q_i}\neq 0$~.

\section{RGE functions}\label{app:RGE}

Eq.~\eqref{eq:acp_zprime_formula}  for $\Delta A_{\text{CP}}^{\text{NP}}$ accounts for the running and mixing of  operators through the functions $r_{1,2}$. The latter can be obtained from the evolution of the Wilson coefficients described in Appendix~\ref{app:A}. 
We obtain
\begin{align}
     r_1(m_c,M_{Z^\prime})&= \frac{R^{-2}}{3\,\sqrt{2}\,G_F\,\lambda_s}~,\\
     r_2(m_c,M_{Z^\prime})&= \frac{2\,R^{1/2}-R^{-1}}{3\,\sqrt{2}\,G_F\,\lambda_s}~,
\end{align}
where 
\begin{align}
    R=\bigg(\frac{\alpha_s^{(4)}(m_b)}{\alpha_s^{(4)}(m_c)}\bigg)^{\frac{12}{25}}\bigg(\frac{\alpha_s^{(5)}(m_t)}{
  \alpha_s^{(5)}(m_b)}\bigg)^{\frac{12}{23}} \bigg(\frac{\alpha_s^{(6)}(M_{Z^\prime})}{\alpha_s^{(6)}(m_t)}\bigg)^{\frac{4}{7}}~.
\end{align}


\end{document}